\documentclass[a4paper,onecolumn,11pt]{quantumarticle}
\pdfoutput=1
\usepackage[utf8]{inputenc}
\usepackage[english]{babel}
\usepackage[T1]{fontenc}
\usepackage{amsmath}
\usepackage{hyperref}
\usepackage{subfigure}
\usepackage{booktabs}
\usepackage{tikz}
\usepackage{lipsum}
\usepackage{txfonts}
\let\mathbb=\varmathbb
\DeclareSymbolFont{letters}{OML}{ztmcm}{m}{it}

\begin{document}

\title{Machine Learning assisted excess noise suppression for continuous-variable quantum key distribution}

\author{Kexin Liang}
\affiliation{Institute for Quantum Information and Technology, School of Information Science and Technology, Northwest University, Xi'an 710127, China.}
\affiliation{State Key Laboratory of Integrated Services Networks (Xidian University), Xi'an, 710071, China.}
\orcid{0000-0003-4749-9323}
\author{Geng Chai}
\orcid{0000-0002-9805-5020}
\thanks{chai.geng@nwu.edu.cn}
\affiliation{Institute for Quantum Information and Technology, School of Information Science and Technology, Northwest University, Xi'an 710127, China.}
\author{Zhengwen Cao}
\affiliation{Institute for Quantum Information and Technology, School of Information Science and Technology, Northwest University, Xi'an 710127, China.}
\affiliation{State Key Laboratory of Integrated Services Networks (Xidian University), Xi'an, 710071, China.}
\thanks{caozhw@nwu.edu.cn}
\author{Qing Wang}
\affiliation{Institute for Quantum Information and Technology, School of Information Science and Technology, Northwest University, Xi'an 710127, China.}
\author{Lei Wang}
\affiliation{Institute for Quantum Information and Technology, School of Information Science and Technology, Northwest University, Xi'an 710127, China.}
\author{Jinye Peng}
\affiliation{Institute for Quantum Information and Technology, School of Information Science and Technology, Northwest University, Xi'an 710127, China.}

\maketitle

\begin{abstract}
Excess noise is a major obstacle to high-performance continuous-variable quantum key distribution (CVQKD), which is mainly derived from the amplitude attenuation and phase fluctuation of quantum signals caused by channel instability. 
Here, an excess noise suppression scheme based on equalization is proposed. In this scheme, the distorted signals can be corrected through equalization assisted by a neural network and pilot tone, relieving the pressure on the post-processing and eliminating the hardware cost. For a free-space channel with more intense fluctuation, a classification algorithm is added to classify the received variables, and then the distinctive equalization correction for different classes is carried out. The experimental results show that the scheme can suppress the excess noise to a lower level, and has a significant performance improvement. Moreover, the scheme also enables the system to cope with strong turbulence. It breaks the bottleneck of long-distance quantum communication and lays a foundation for the large-scale application of CVQKD.
\end{abstract}

\section{Introduction}
The development of digitalization and intelligentization in modern society is based on massive information exchange. Therefore, how to realize the secure transmission of information is an important development direction in modern communication. 
Quantum key distribution (QKD)  \cite{r1,r1c,r1d,r3,zongshu1} combined with the One-time pad (OTP) can achieve theoretically unconditional secure communication. It can be implemented in the discrete-variable scheme and continuous-variable scheme (CVQKD) \cite{CV1, CV2},  and the latter has the advantages of a high secret key rate (SKR) and low cost. Long-distance transmission \cite{202.81} and high SKR \cite{highSKR1,highSKR2} are the two critical goals of CVQKD, which are limited by the practical transmission loss of the quantum channel, excess noise, and reconciliation efficiency \cite{ r9,r10}. Therefore, maintaining high reconciliation efficiency \cite{recon1,recon2} and low excess noise is a major obstacle to the realization of remote CVQKD. 

Excess noise refers to the sum of the variances of all possible noise sources in Alice's system (imperfect modulation), channel transmission (phase fluctuations), and Bob's system (imperfect detection, electronic noise), which results in a serious system security threats \cite{excess1,excess2, twotimes, l4, l5}. 
There are two solutions to eliminate its influence. One is to improve the tolerance of the system, such as the two-way protocol \cite{tolerance1} and the phase noise model proposed in \cite{tolerance2}. The other is to suppress excess noise at a lower level which is mainly realized by tracking and compensating for phase noise \cite{track1,track2,track3,compensation1,compensation2}. A phase estimation protocol based on the theoretical security and Bayes’ theorem is studied in \cite{track1}, which can achieve a well-motivated confidence interval of the estimated eigenphase without the strong reference pulse propagation.  In the work \cite{track2,track3}, the fast and slow phase drift can be both estimated by using the improved vector Kalman filter carrier phase estimation algorithm, and thus the phase estimation error can be tracked in real-time and be almost approximate to the theoretical mean square error limit. An implementation of a machine learning framework based on an unscented Kalman filter is explored in \cite{compensation2} for estimation of phase noise, enabling CVQKD systems with low hardware complexity which can work on diverse transmission lines. The above studies have achieved a great excess noise suppression effect, but they are all from the perspective of improving the accuracy and real-time performance of the phase noise estimation and compensation, not from its source. The free-space channel has greater fluctuation due to the atmospheric turbulence effect, research on free-space excess noise suppression is also in progress \cite{fs1,fspc1,fspc2}.

This work proposes an excess noise suppression scheme based on equalization assisted by machine learning algorithms for CVQKD.  Channel equalization is an anti-fading measure taken to improve the transmission performance of communication systems in fluctuating channels, therefore, equalization can fundamentally reduce the excess noise caused by channel fluctuation. This scheme obtains the correction coefficients through the neural network acting on pilot tone, makes parameters estimation to judge the transmission safety, then applies the trained correction coefficients to the signal following pilot tone at the same stage to reduce the impact of channel fluctuation on its quality, so as to suppress excess noise and improve system performance.
This correction performance is verified experimentally in a Gaussian modulated coherent state (GMCS) CVQKD \cite{r7,r8} under a 10km fiber channel, which achieves a more stable and lower excess noise. 
A free-space communication process experiences turbulences of different intensities, thus, the quality of all received variables varies greatly, which affects the training effect of the correction coefficients. A K-Nearest Neighbor (KNN) classification algorithm is employed to divide these received variables into three classes according to their quality. For each class, distinctive correction coefficients are trained to better complete the signal correction and improve the overall performance of the system. Compared with the unclassified equalization scheme, the classified scheme has a better correction effect, especially in medium and strong turbulence, thus, this scheme enables the CVQKD system to curb the negative impact of transmission fluctuation on the system.
\section{CVQKD system with equalization}
\begin{figure*}[!htbp]\center
\centering
\resizebox{14cm}{!}{
\includegraphics{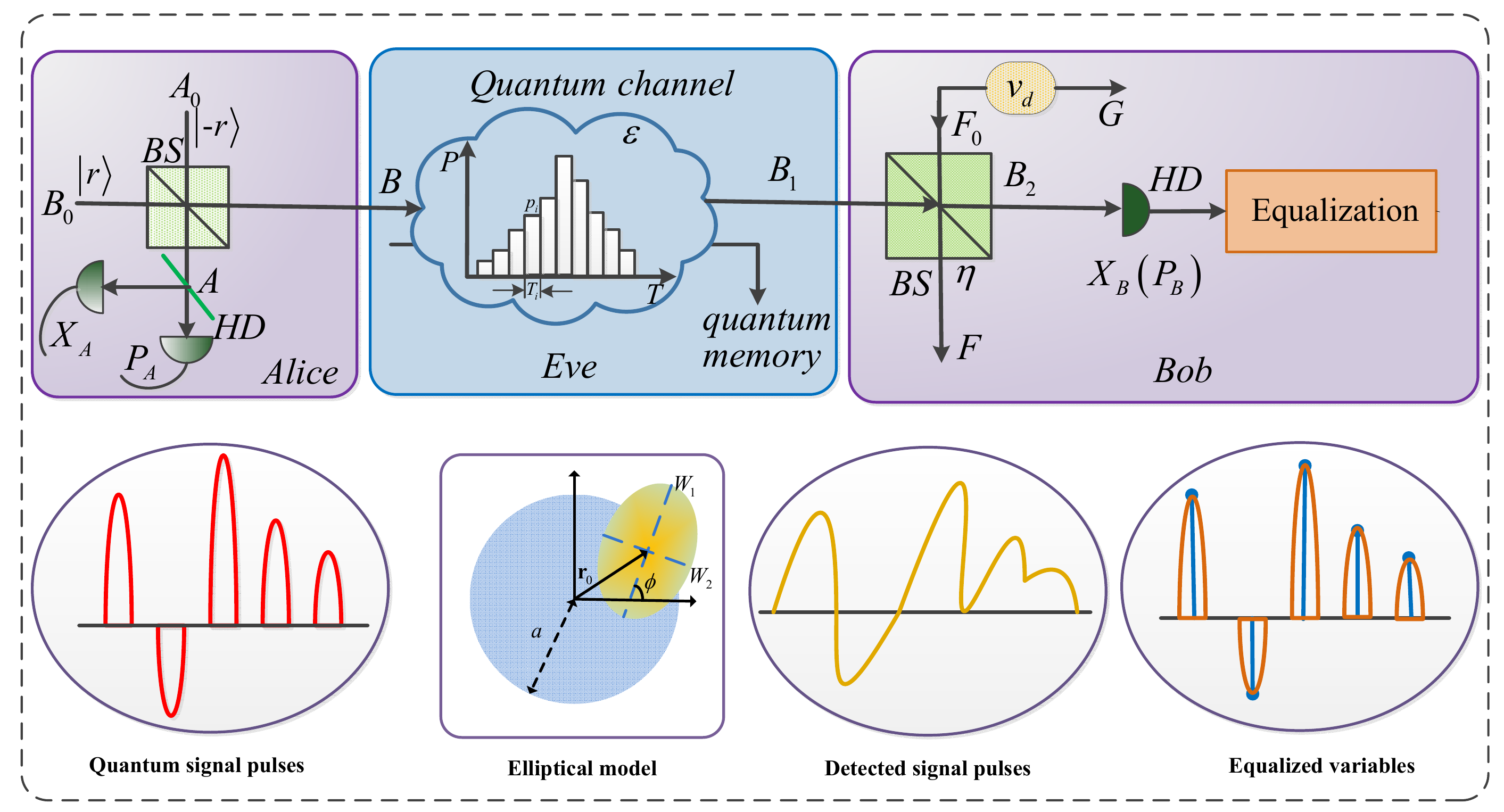}}
\caption{\textbf{EB model of CVQKD system with equalization.} (BS: Beam splitter, HD: Homodyne detection, $\eta $: detection efficiency, $T$: transmission rate, $\varepsilon$: excess noise. The ellipse model is used to describe the beam wangdering caused by the beam passing through the free-space channel, where $(W_{1},W_{2},\phi  )$) indicates the direction and size of the beam, $a$ is aperture radius, and $r_{0} $ is the center of beam.}
\label{fig:9}
\end{figure*}

As shown in Fig. \ref{fig:9},  Alice modulates the quantum signals and then dispatches them to Bob via a quantum channel whose feature is transmittance distribution $P(T)$. After taking over the quantum signal, Bob conducts the coherent detection of the received signals and acquires the raw key variables. A practical detector is featured by an efficiency $\eta $ and a noise $\upsilon _{el} $ on account of detector electronics. Detected signals deviate due to the random fluctuation of the quantum channel. Before Bob extracts secret keys, he first preprocesses the received variables with equalization to make the processed received variables close to the ideal variables as possible, thus, there is only atmospheric inherent attenuation in the quantum channel and no atmospheric turbulence so as to make up for the bad impact of atmospheric turbulence on the transmission signal.

Specifically,  Alice employs two single-mode squeezed vacuum states $\left|  \nu   \right \rangle $ and $\left| - \nu   \right \rangle $ to prepare a two-mode squeezed $\left | \psi  \right \rangle _{AB} $. Then she measures one half of it randomly and sends another one to Bob through a quantum channel.  
The state  $\left | \psi  \right \rangle _{AB} $  after transmission through free-space channel is described as
\begin{eqnarray}
\left | \psi  \right \rangle _{AB_{1} } =\left \{ \mathbf{I} \otimes U_{trans(L)}  \right \} \left | \psi  \right \rangle _{AB },
\end{eqnarray}
where $U_{trans(L)}$ indicates the effect of channel on quantum state. $\mathbf{I}$ means that the first module is saved at Alice and is not transmitted through the channel, so as to maintain its original state. The second mode undergoes channel transmission and is affected by $U_{trans(L)}$.
The covariance matrix $\gamma _{AB_{1} } $ of $\left | \psi  \right \rangle _{AB_{1} } $  is written as
\begin{equation}
\begin{split}
&\gamma _{AB_{1} } =\left ( \mathbf{I} \otimes U_{trans(L)}  \right ) ^{T} \gamma _{AB } \left ( \mathbf{I} \otimes U_{trans(L)}  \right )=\\
&\begin{pmatrix}
  V&  0& \Lambda cos(\Delta \varphi)  & -\Lambda sin(\Delta \varphi )\\
  0& V &-\Lambda sin(\Delta \varphi ) &-\Lambda cos(\Delta \varphi )  \\
\Lambda cos(\Delta \varphi )& -\Lambda sin(\Delta \varphi )& T \left (V+\frac{1}{T}-1+\varepsilon    \right )  & 0\\
  -\Lambda sin(\Delta \varphi )& -\Lambda cos(\Delta \varphi )  & 0 &T \left (V+\frac{1}{T}-1+\varepsilon    \right )
\end{pmatrix}
\end{split},
\end{equation}
where $\Lambda =\sqrt {A_{\alpha }^{eq}}\sqrt{T^{th}} \sqrt{V^{2}-1 }$, $T^{th} $ is theoretical channel transmission, $A_{\alpha }$, $\Delta \varphi $, $T$ and $\varepsilon$ are amplitude attenuation, phase drift, and practical channel transmittance and excess noise. $V$ is variance of $\left | \psi  \right \rangle _{AB }$.

When the sampled variable is corrected by the equalization, it is equivalent to correcting the quantum signal to make it closer to the ideal data. 
\begin{eqnarray}
\left | \psi  \right \rangle _{AB_{2} }  =  U_{eq} \left | \psi  \right \rangle  _{AB_{1} }  =  \left \{ \mathbf{I}\otimes  U_{eq}U_{trans}(L) \right \}\left | \psi  \right \rangle _{AB},
\end{eqnarray}
where $U_{eq}$ indicates the correction effect on quantum signal correction.
Thus, the covariance matrix $\gamma _{AB_{2} } $ of $\left | \psi  \right \rangle _{AB_{2} } $  is rewritten as
\begin{equation*}
\footnotesize{
\begin{split}
&\gamma _{AB_{2} } =\left ( \mathbf{I} \otimes U_{eq)}  \right ) ^{T} \gamma _{AB_{1}  } \left ( \mathbf{I} \otimes U_{eq}  \right )=\\
&\begin{pmatrix}
  V&  0&\frac{ \Lambda} { \sqrt{A_{\alpha }^{eq}}}cos  (\Delta \varphi-\Delta \varphi^{eq} ) & -\frac{ \Lambda} { \sqrt{A_{\alpha }^{eq}}}sin  (\Delta \varphi-\Delta \varphi^{eq} )\\
  0& V &-\frac{ \Lambda} { \sqrt{A_{\alpha }^{eq}}}sin (\Delta \varphi-\Delta \varphi^{eq} ) &-\frac{ \Lambda} { \sqrt{A_{\alpha }^{eq}}}cos (\Delta \varphi-\Delta \varphi^{eq} )  \\
\frac{ \Lambda} { \sqrt{A_{\alpha }^{eq}}}cos(\Delta \varphi-\Delta \varphi^{eq} )  & -\frac{ \Lambda} { \sqrt{A_{\alpha }^{eq}}}sin(\Delta \varphi-\Delta \varphi^{eq} )  & T^{eq} \left (V+\frac{1}{T^{eq}}-1+\varepsilon^{eq}    \right )  & 0\\
  -\frac{ \Lambda} { \sqrt{A_{\alpha }^{eq}}}sin(\Delta \varphi-\Delta \varphi^{eq} ) & -\frac{ \Lambda} { \sqrt{A_{\alpha }^{eq}}}cos(\Delta \varphi-\Delta \varphi^{eq} )  & 0 &T^{eq} \left (V+\frac{1}{T^{eq}}-1+\varepsilon^{eq}    \right )
\end{pmatrix}
\end{split}},
\end{equation*}
where $A_{\alpha }^{eq}$, $\Delta \varphi^{eq}$, $T^{eq}$ and $\varepsilon^{eq}$  are amplitude attenuation, phase drift, channel transmittance and excess noise with equalization correction. Under ideal conditions, $\frac{A_{\alpha }^{eq}}{A_{\alpha }} =1$ and $\Delta \varphi^{eq}=\Delta \varphi$, which means perfect correction.
\section{Experimental vertification under fiber channel}
\subsection{Establishment of dataset}
\begin{figure}[!htbp]\center
\centering
\resizebox{14cm}{!}{
\includegraphics{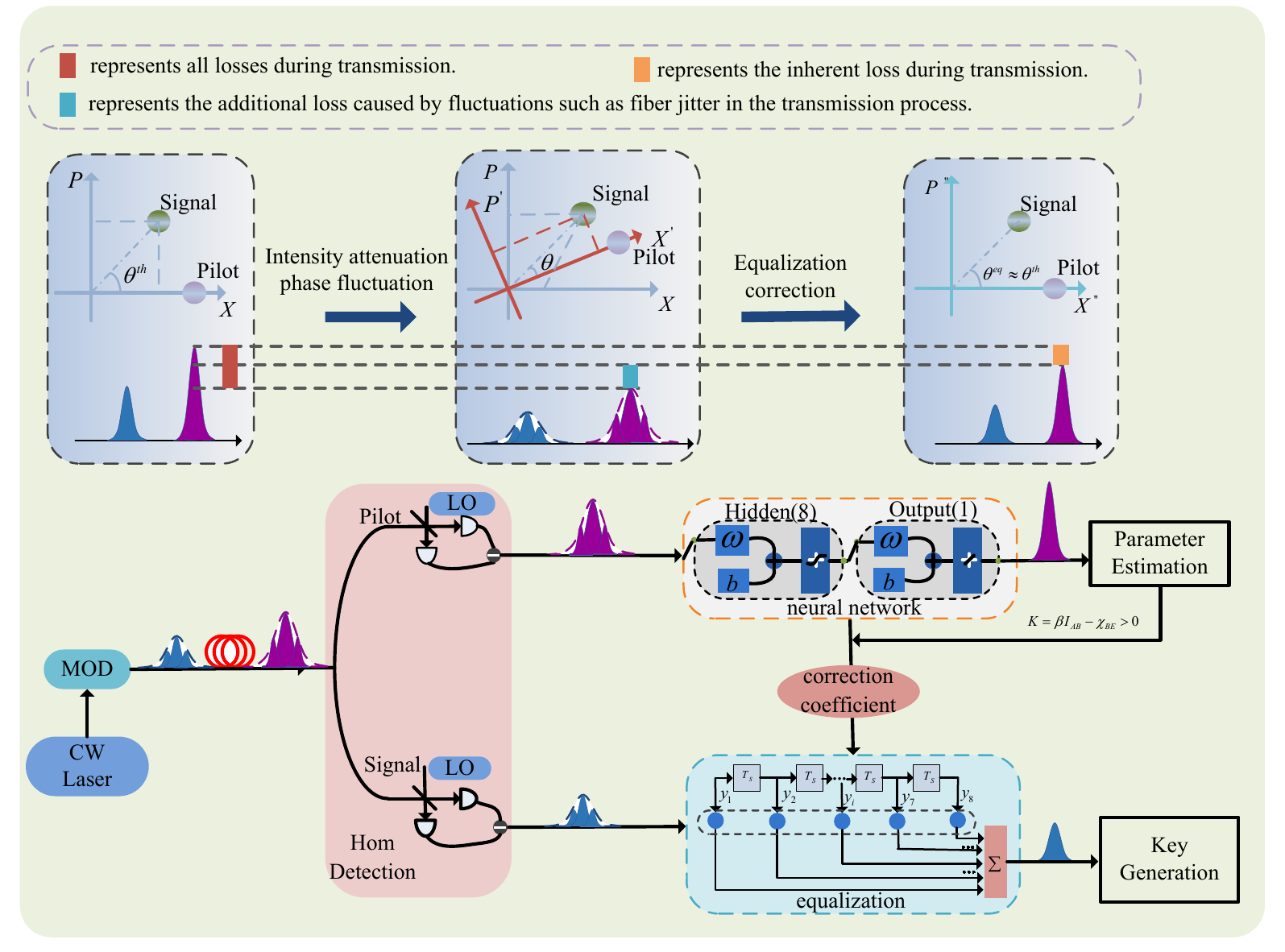}}
\caption{\textbf{Experimental setup.} (CW: continuous wave, LO:local oscillator, MOD: modulation module.  $\theta^{th}$, $\theta$, and $\theta^{eq}$ represent phase relationship between signal and pilot at the sender, the receiver, and after correction, respectively. $y_{i} (i=1, 2, ..., 8)$ is sampled point of received signal pulse, $T_{s} $ is sampling interval, $\omega $ and $b$ are the weight and bias of the neural network. )}
\label{fig:10}
\end{figure}

The experimental setup is shown in Fig. \ref{fig:10}.
On Alice's side, she prepares two independent Gaussian random variables $X_{A} $ and $P_{A}$ which obey the same zero-centered Gaussian distribution $\mathcal{N}  (0,V_{A} )$, where $V_{A}$ is the modulation variance. Then she sends modulated coherent state $\left | \alpha _{s}   \right \rangle =\left |X_{A} +iP_{A} \right \rangle  $ as quantum signal, and sends another classical coherent state $\left | \alpha _{P}   \right \rangle =\left |X^{P} _{A} +iP^{P}_{A} \right \rangle $ as pilot tone in the next time bin to Bob where $(X^{P} _{A},P^{P} _{A})$ is publicly known \cite{constant2,constant1}. 
Repeating this process many times, interleaved signal pulses and pilot tone are simultaneously transmitted from Alice to Bob. After the signal is transmitted through the fluctuating channel, it will produce intensity attenuation and phase fluctuation for quantum signals.  On Bob's side, he employs a local oscillator (LO) to perform homodyne detection with a pilot tone pulse and with a signal pulse, respectively.  After the pilot tone is corrected by the neural network, its parameters $(T^{eq} ,\varepsilon^{eq} )$ are estimated. If the transmission is evaluated as safe, the trained correction coefficient can also be used for the signal pulse at the same stage, which is helpful for the next key generation. 

\subsection{Performance analysis}
The theoretical transmittance $T^{th}$ is mainly related to the transmission distance $L$. However, the transmission fluctuation produces more excess noise in the system, which makes the practical transmittance $T$ less than $T^{th}$. In order to make practical channel transmission performance as close to ideal channel transmission performance as possible at the same distance, it is necessary to suppress the negative impact of the channel fluctuation on the signal.  $T^{th}$ can be obtained by 
\begin{eqnarray}
T^{th}=10^{-\alpha _{f}L/10 },
\end{eqnarray}
where $\alpha _{f}(dB/km)$ is the attenuation coefficient.
Using the received variables of the known pilot tone, the neural network can be applied to predict the ideal received values quickly and accurately, and $T^{th}$ provides a target for it.
The purpose of using neural network at Bob side is
\begin{eqnarray}
f(\mathbf{y}  )  =  \mathbf{\omega } ^{T} \mathbf{y}_{i} +b  \approx {y}' = t^{th}x,
\end{eqnarray}
where $\mathbf{y}$ is an $8 \times 1$ vector representing eight points on a pilot tone pulse acquired by the oscilloscope at one time, $ \mathbf{\omega }$ is the correction coefficient vector trained by the neural network,  $f(\mathbf{y}  )$  represents the corrected value, $ {y}' $ is the targeted  received variable, $t^{th}$ is the parameter in the linear channel model that is provided by $T^{th}$  $(t^{th} =\sqrt{\eta T^{th} }$, $\eta$ is detection efficiency), and $x$ is Alice modulated variable, $b$ is bias of neural network. 
\begin{figure}[!htbp]\center
\centering
\resizebox{14cm}{!}{
\includegraphics{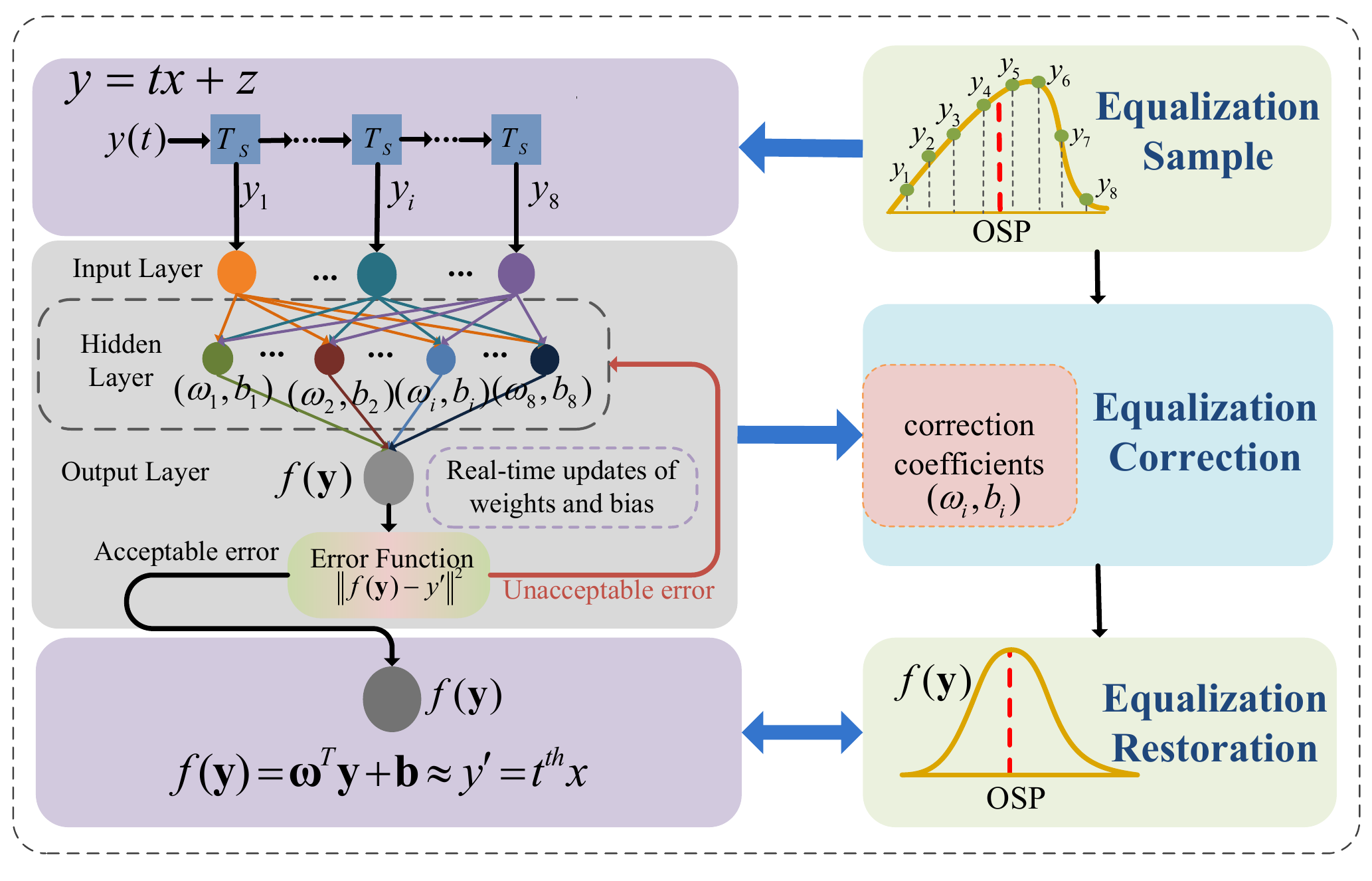}}
\caption{\textbf{Equalization correction via neural network.} (
 OSP: optimum sampling point, $T_{s}$ is time delay during sampling, $\omega_{i}$ and $b_{i}$ are weights and bias trained by neural network, $ t=\sqrt{\eta T} $  represents practical channel state, $ t^{th}=\sqrt{\eta T^{th}} $  represents ideal channel state, ${y}_{i}(i=1,2...,8)$ is sampling points on a pilot tone pulse acquired by the oscilloscope at one time, ${y}'$ refers to targeted received variable, $z$ represents Gaussian noise, and $x$ is Alice modulated variable. ) }
\label{fig:0}
\end{figure}

\begin{figure*}[htbp]
\begin{center}
\subfigure[\textbf{The difference between $x$ and $y$ versus the difference between $x$ and $f(\mathbf{y}) $.}]{
\includegraphics[width=0.45\linewidth]{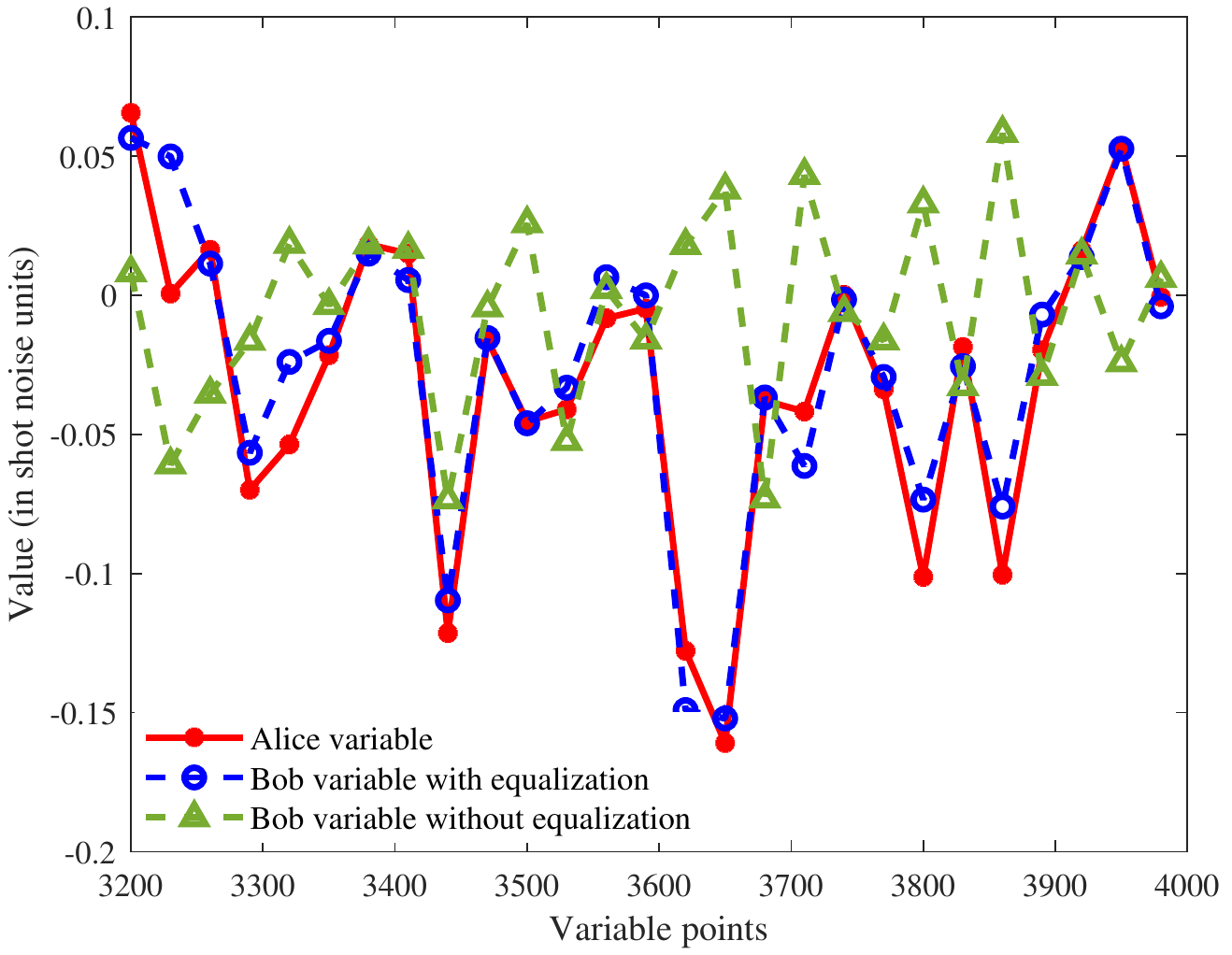}
\label{fig:11}
}
\subfigure[ \textbf{Excess noise suppression by equalization.} ]{
\includegraphics[width=0.45\linewidth]{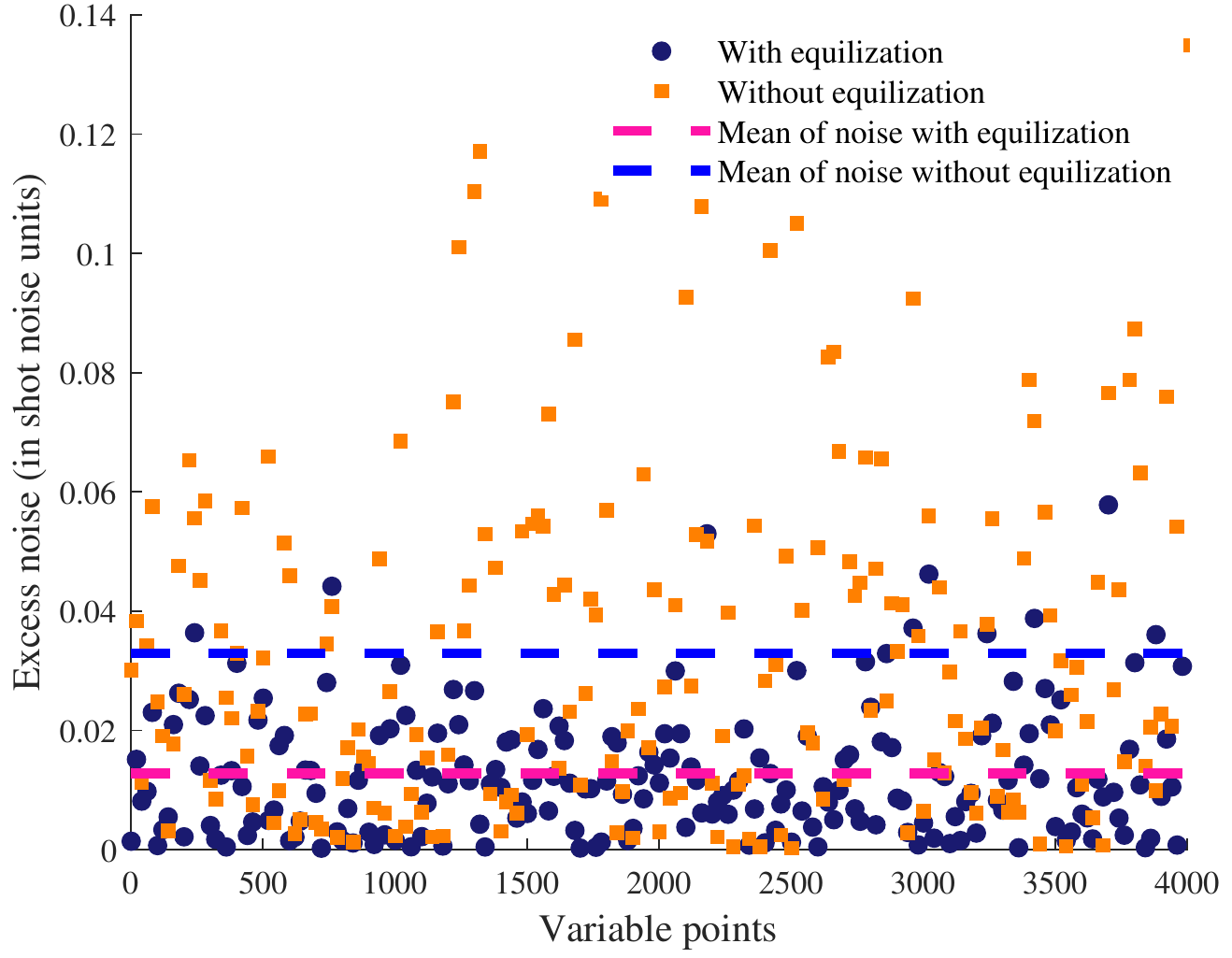}
\label{fig:12}
}
\caption{\textbf{Correction of received variable and excess noise suppression by equalization.} (In Fig. \ref{fig:11}, the solid line with red filled circles, the dotted line with the blue hollow circles, and  the dotted line with the green triangles represent Alice modulated variable $x$, equalized variable $f(\mathbf{y})$, and practically received variable $y$, respectively. In Fig. \ref{fig:12}, blue circles and orange squares present the excess noise of the system with and without equalization, respectively, and the below-dotted line and the upper-dotted line represent their mean values.)}  
\label{fig:1}
\end{center}
\end{figure*}

The principle of equalization correction via a neural network is shown in Fig. \ref{fig:0}, where the neural network has 8 input neurons, 1 output neuron, and one hidden layer. A received pilot tone pulse is sampled into eight points $(y_{1}, y_{2},...,y_{8})$, which are the input of the neural network. After training the neural network, the predicted received variable $f(\mathbf{y})$ is obtained. By comparing with the expected variable ${y}'$, if the error is within a reasonable range, the neural network is applied to the signal pulse following the pilot tone at the same stage. If the error exceeds the acceptable range, the neural network will be retrained, and the weight and bias $(\omega _{i},b_{i}  )$ will be adjusted, so that the equalization system has a certain adaptive ability. The results of signal correction and suppression of system excess noise by equalization are shown in Fig. \ref{fig:1}. 

Let Alice modulated variable be $x$ and Bob received variable be $y$ (without equalization) and $f(\mathbf{y})  $ (with equalization). It can be seen that  $f(\mathbf{y}) $ is closer to $x$ in Fig. \ref{fig:11}. The correlation $\gamma_{xf(\mathbf{y}) }$ between $f(\mathbf{y}) $ and $x$ enhances than that $\gamma_{xy } $ between $y$ and $x$ according to parameter estimation.
In Fig. \ref{fig:12}, the excess noise with equalization $\varepsilon^{eq}$ and without equalization  $\varepsilon $ are calculated, respectively, and the former features a more concentrated distribution and a lower mean value. 

\begin{figure}[!htbp]\center
\centering
\resizebox{9cm}{!}{
\includegraphics{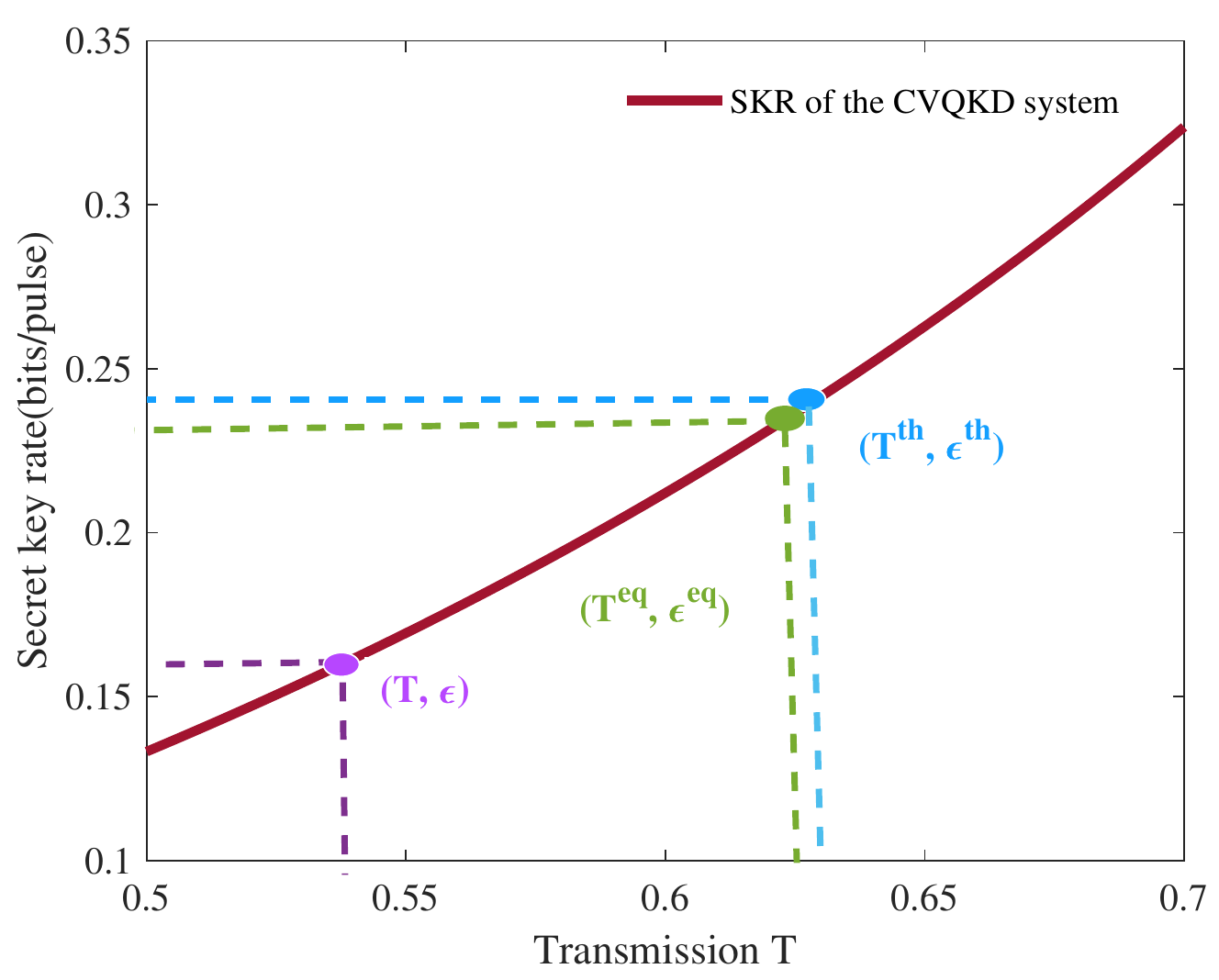}}
\caption{\textbf{System performance enhancement of the proposed equalization scheme.} (From left to right, the first purple dot, the second green dot, and the last blue dot describe the system performance without equalization, with equalization, and under ideal conditions, respectively. The red line is the SKR under different transmission environments.)}
\label{fig:2}
\end{figure}

Based on the $(x,y)$ and $(x, f(\mathbf{y}) )$, the practical system parameters $(T,\varepsilon )$, the equalized system parameters $(T^{eq} ,\varepsilon^{eq} )$, and theoretical system parameters $(T^{th},\varepsilon^{th} )$ are calculated in Tab. \ref{tab:01}. The transmission loss $(1-T^{eq}) $ between $f(\mathbf{y}) $ and $x$  is mainly caused by the inevitable loss of the channel. And the improvement of system performance is shown in Fig. \ref{fig:2}. It can be seen that equalization is of great help to improve the system performance, which is close to the theoretical SKR.  In the farthest CVQKD experiment using phase compensation \cite{202.81}, the excess noise suppression rate is 78\%. This proposed scheme achieves 72\% without additional phase compensation software and hardware.
\begin{table}[!h]
\caption{Comparisons of transmission and excess noise.}
\centering
\label{tab:01}
\begin{tabular}{cccc}
   \toprule[1.2pt]
               & without equalization & with equalization & theoretical value  \\
   \midrule[1pt]
   transmission & $T=0.5412$ & $T^{eq}=0.6261$ & $T^{th}=0.6310 $\\
   excess noise & $\varepsilon =0.0429$  &$\varepsilon^{eq}=0.0128$ & $\varepsilon^{th}=0.01$ \\
   \bottomrule[1.2pt]
\end{tabular}
\end{table}


\section{Simulation analysis under free-space channel}

\subsection{Establish data set through Monte Carlo}
\begin{figure}[!htbp]\center
\centering
\resizebox{10cm}{!}{
\includegraphics{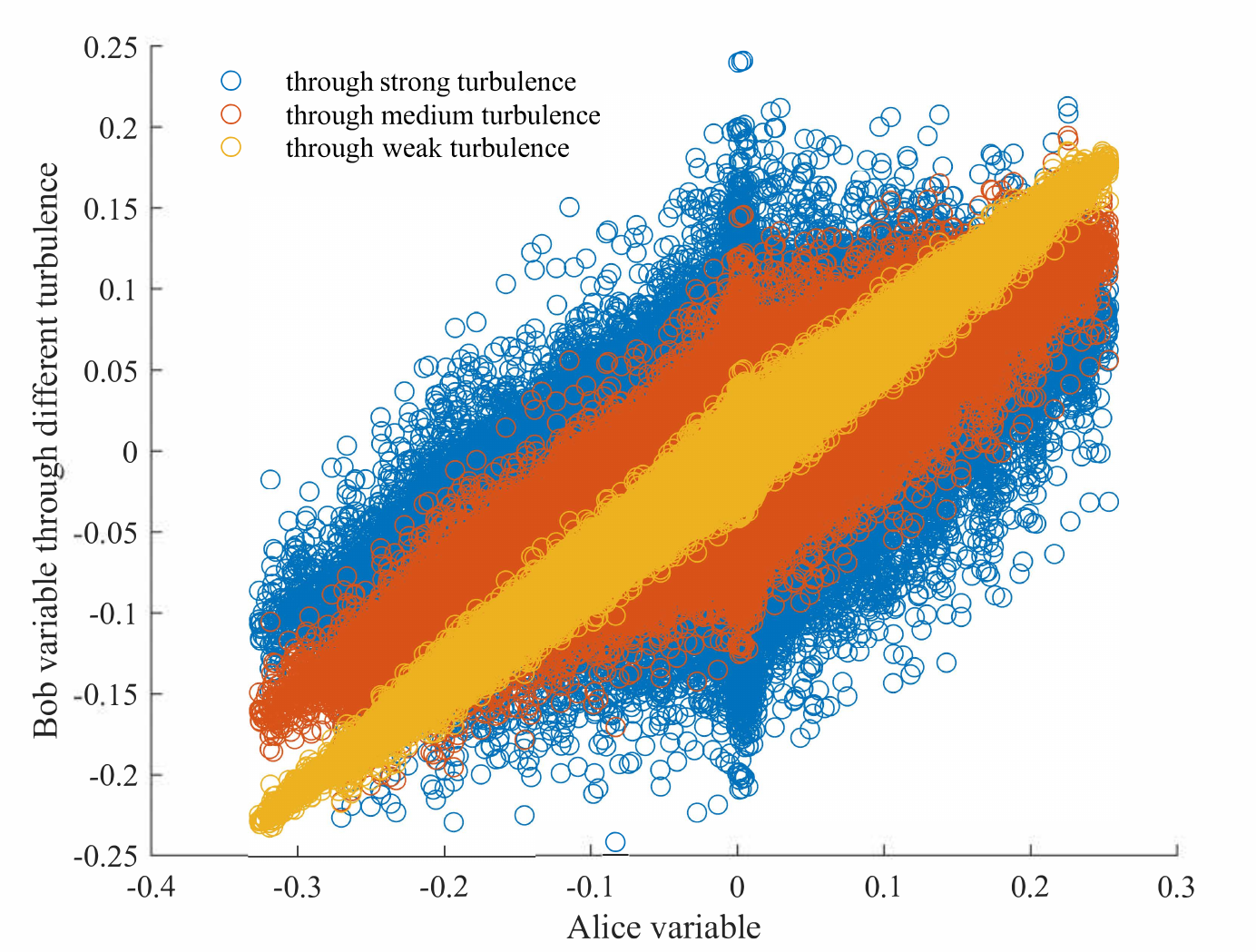}}
\caption{\textbf{Correlation between the received variables constructed under different turbulence and Alice modulation variables.} (Yellow circles, orange circles, and blue circles are the received variables through weak turbulence, medium turbulence, and strong turbulence.)}
\label{fig:3}
\end{figure}

In the atmospheric channel, the theoretical transmittance $T^{fs}$ can be obtained according to Lambert's law \cite{zong},
\begin{eqnarray}
T^{fs}=\mathrm{exp} (-\alpha _{\lambda } L),
\end{eqnarray}
where $\alpha _{\lambda }$  stands for the wavelength-dependent coefficient of atmospheric decay that changes greatly under different weather conditions. 
The empirical formula of $\alpha _{\lambda } $ and air visibility $V$ is \cite{visibility} 
\begin{eqnarray}
\alpha _{\lambda } =\frac{3.91}{V} \times \left ( \frac{\lambda }{0.55}  \right ) ^{-q} ,
\end{eqnarray}
where $q$ is a constant. Thus, the typical value of  $\alpha _{\lambda } $ under different weather conditions can be obtained. 
Three simulated received variables $y_{weak} $, $y_{medium} $ and $y_{strong} $ under weak turbulence, medium turbulence, and strong turbulence are constructed according to different $\alpha _{\lambda } $, respectively, as shown in Tab. \ref{tab:02}. And statistical relationships between simulated received variables and Alice modulated variable $(x, y_{weak})$, $(x, y_{medium})$, and $(x, y_{strong})$ are shown in Fig. \ref{fig:3}.
\begin{table*}[!h]
\caption{$\alpha _{\lambda }$, corresponding transimission and received variables under different  turbulences.}
\centering
\label{tab:02}
\begin{tabular}{cccc}
   \toprule[1.2pt] 
               & weak turbulence& medium turbulence &  strong turbulence\\
               &(sunny) &  (rainy) &   (stormy)\\
   \midrule[1pt]
   $\alpha _{\lambda }(dB/km) $ & 2 &15 & 30  \\
  $T^{fs}$ & 0.9703 &0.5003 &0.2352 \\
reveived variable & $y_{weak} $ & $y_{medium} $ &$y_{strong} $\\
   \bottomrule[1.2pt]
\end{tabular}
\end{table*}
When passing through weak turbulence, the relationship of $(x,y_{weak})$ shows good linear characteristics. In this turbulence condition, the attenuation degree is less than that of a fiber channel at the same distance, which also shows the advantages of a free-space channel over a fiber channel and becomes the primary choice for long-distance secret key distribution. 
In medium turbulence case, the linear relationship of $(x,y_{medium})$ decreases. And the linear relationship of $(x,y_{strong})$ under strong turbulence is the worst, therefore, security and long-distance secret key distribution under medium-to-strong turbulence are the bottlenecks that need to be broken through. 

\subsection{Performance analysis}
However, it can also be seen that even though it passes through the strong turbulence with the worst transmission characteristics, some of the variables received by Bob still belong to the part with excellent quality (the part overlapping with the yellow ones), and some belong to the part with ordinary quality (the part overlapping with the orange ones), and the rest belong to the part with bad quality. Therefore, in order to correct variables more accurately, we classify them according to the quality of received variables and then equalize the classified variables with high consistency.

\begin{figure}[!htbp]\center
\centering
\resizebox{10cm}{!}{
\includegraphics{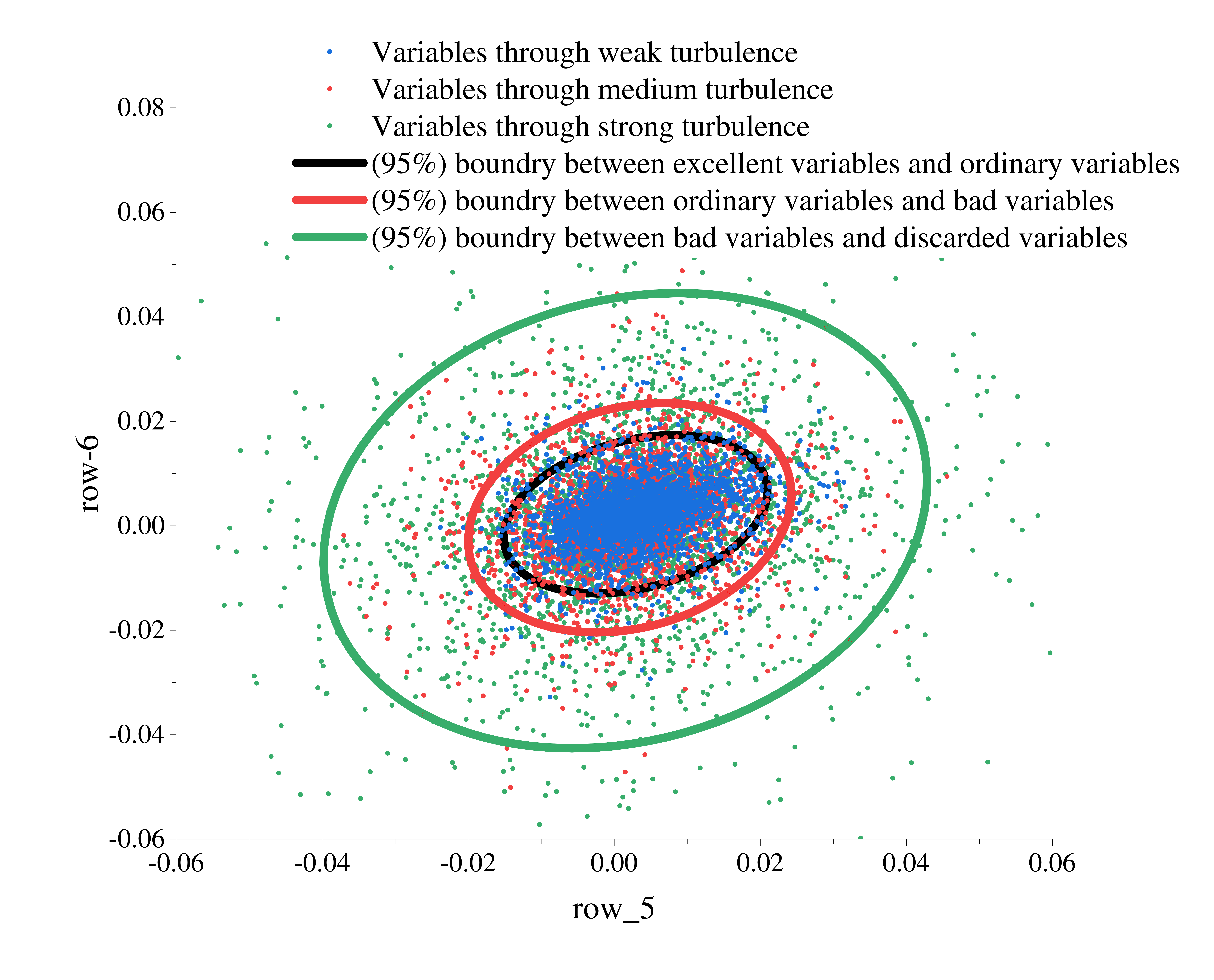}}
\caption{\textbf{Determine the classification labels according to the quality of the received variables. } (Blue dots, orange dots, and green dots represent received variables through weak turbulence, medium turbulence, and strong turbulence, respectively. Variables in the black ellipse, between the black ellipse and the red ellipse, between the green ellipse and the red ellipse, and outside the green ellipse are labeled as excellent variables, ordinary variables,  bad variables, and discarded variables.) }
\label{fig:4}
\end{figure}

First, as shown in Fig. \ref{fig:4}, the training labels are determined according to the ellipse fitting model \cite{qqq}, where the variables are labeled according to these boundaries. The variables within the blue ellipse are considered excellent variables, which are most suitable for extracting secret keys. the variables between the blue ellipse and the red ellipse are considered ordinary variables, the variables between the red ellipse and the green ellipse are considered bad quality variables, which occupy the largest proportion and have the most important impact on system performance, and the variables beyond the green circle are considered variables that need to be discarded.

\begin{figure*}[htbp]
\begin{center}
\subfigure[Confusion Matrix on training set.]{
\includegraphics[width=0.5\linewidth]{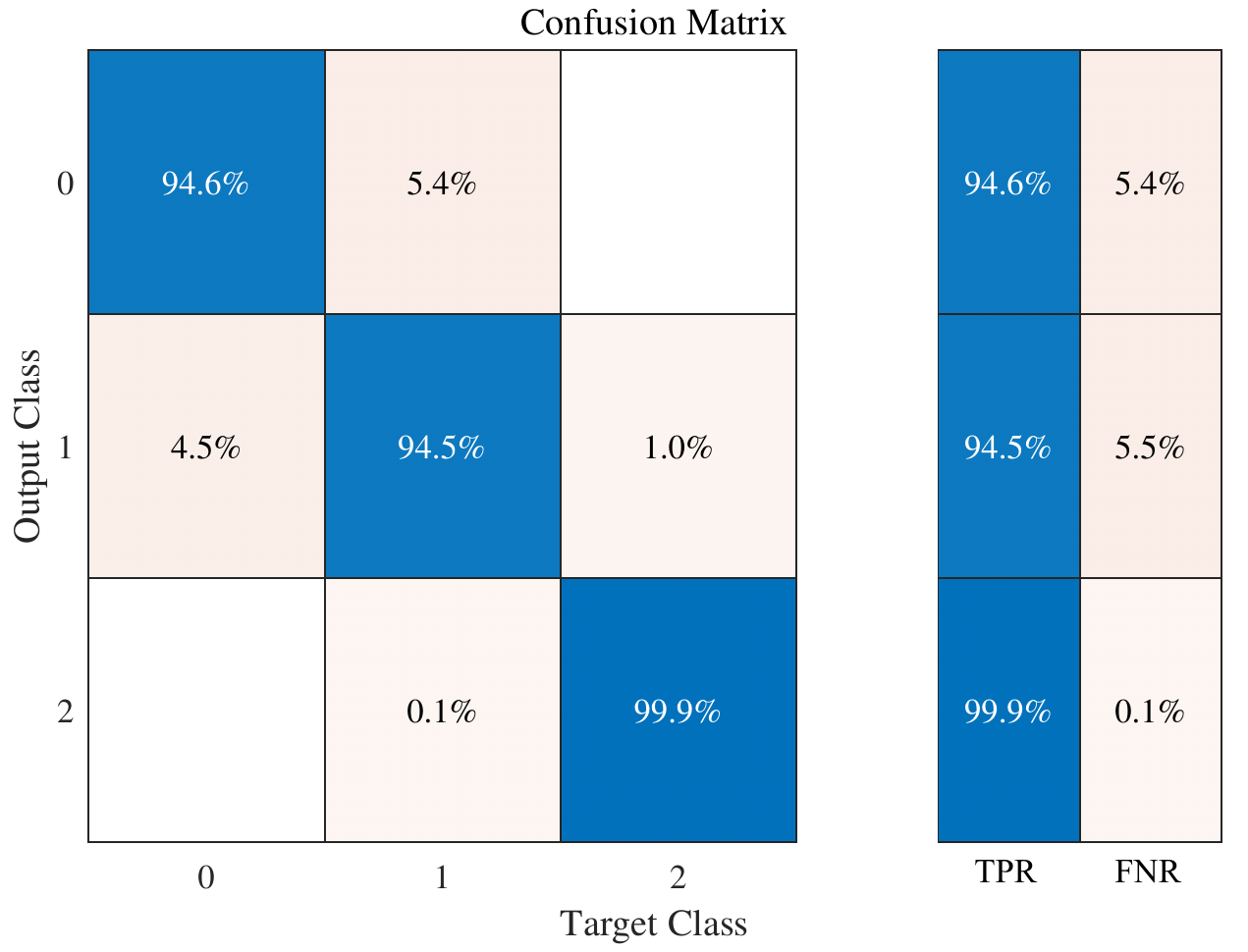}
\label{fig:51}

}
\subfigure[ROC curve on training set.]{
\includegraphics[width=0.45\linewidth]{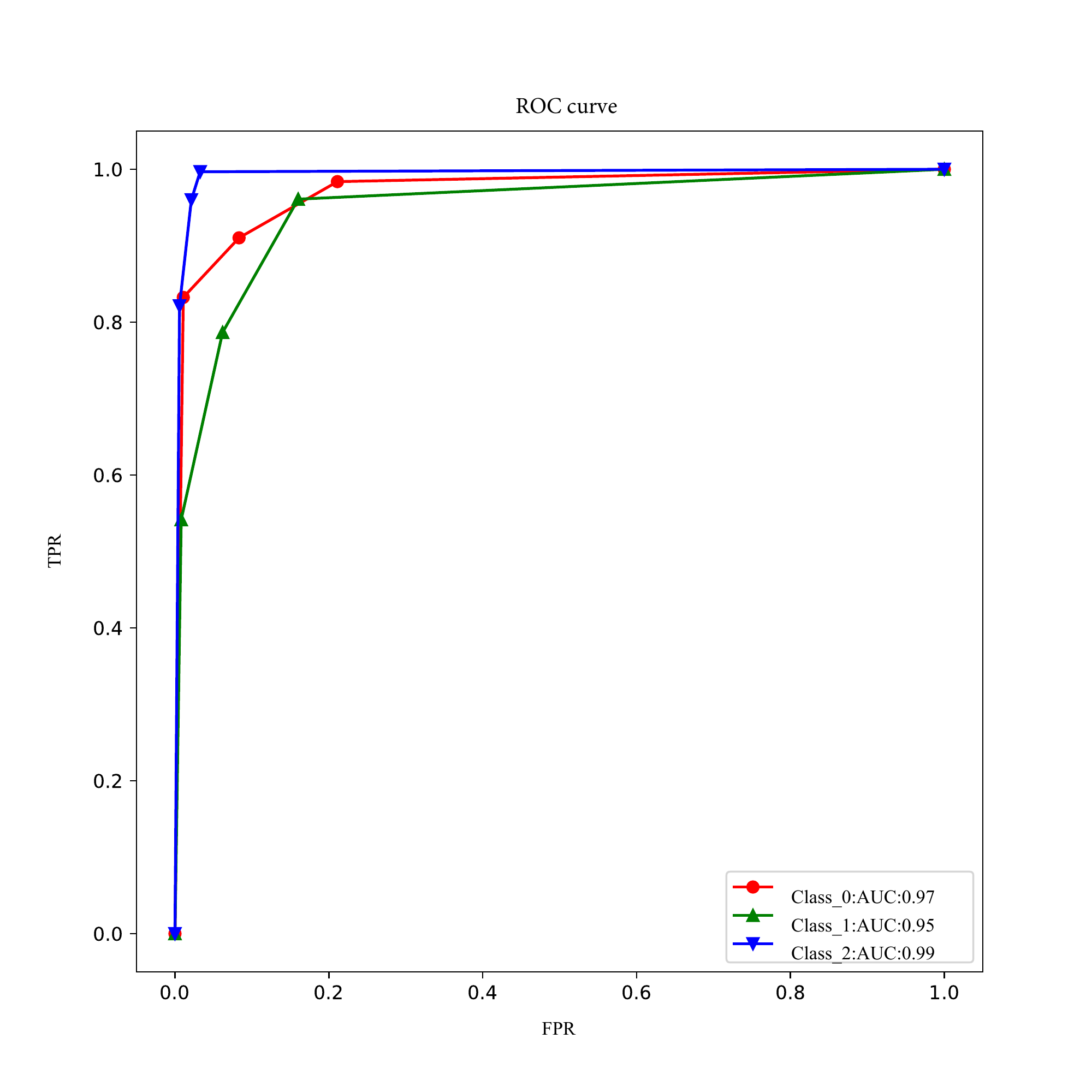}
\label{fig:52}
}
\subfigure[Confusion Matrix on test set.]{
\includegraphics[width=0.5\linewidth]{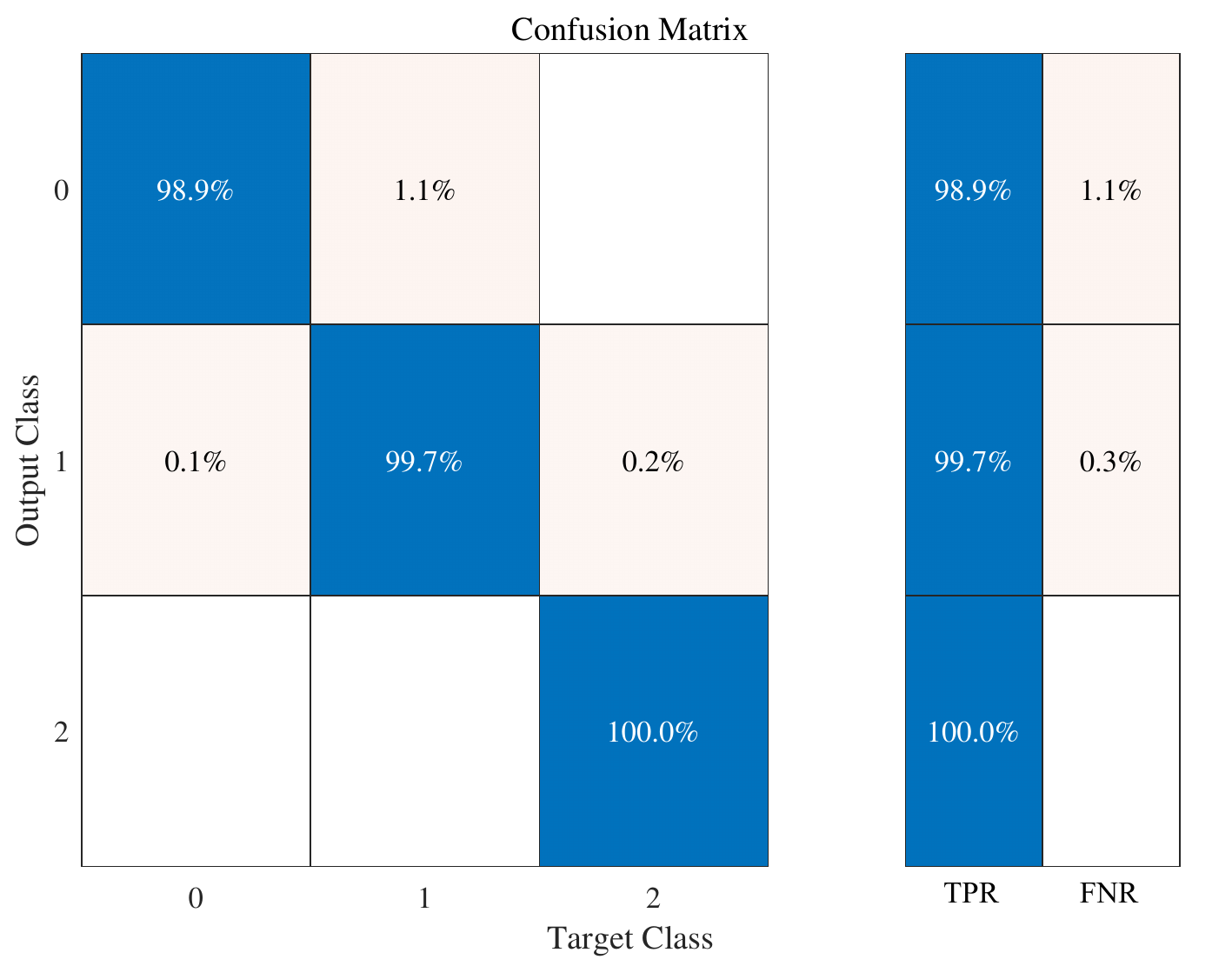}
\label{fig:53}
}
\subfigure[ROC curve on test set.]{
\includegraphics[width=0.45\linewidth]{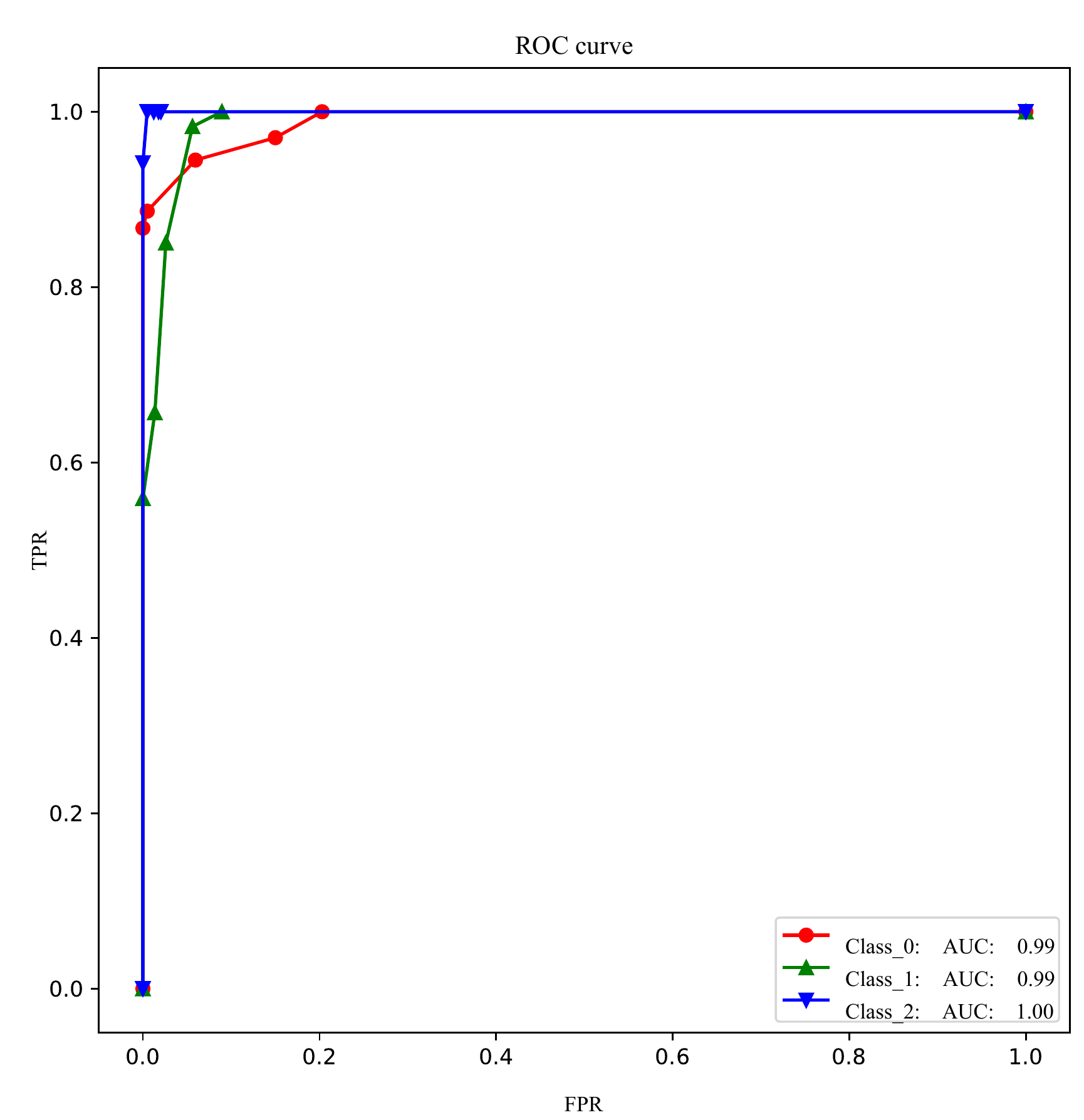}
\label{fig:54}
}
\caption{\textbf{Performance of classifier on the training set and test set.} (TPR: True Positive Rate, FNR: False Negative Rate, ROC (receiver operating characteristic) Curve's abscissa is the false positive rate (FPR) and its ordinate is TPR. The closer the ROC curve is to the upper left corner, the higher the accuracy of the test. AUC (Area Under Curve) is defined as the area surrounded by the coordinate axis under the ROC curve. The closer the AUC is to 1.0, the higher the authenticity of the detection method; Class 0: excellent variables, Class 1: ordinary variables, Class 2: bad variables.)}  
\label{fig:5}
\end{center}
\end{figure*}

Second, train the KNN classifier. KNN is considered to be one of the simplest classification algorithms and the most commonly used classification algorithms. The performances of the classifier on training set and test set are shown in Fig. \ref{fig:5}, and the accuracy is 96.8\% (K=5).
According to the above indicators (Accuracy, TPR, FNR, ROC Curve, and AUC) of the training set and test set, the classifier is trustworthy.

\begin{figure*}[htbp]
\begin{center}
\subfigure[Correction effect under weak turbulent channel.]{
\includegraphics[width=0.43\linewidth]{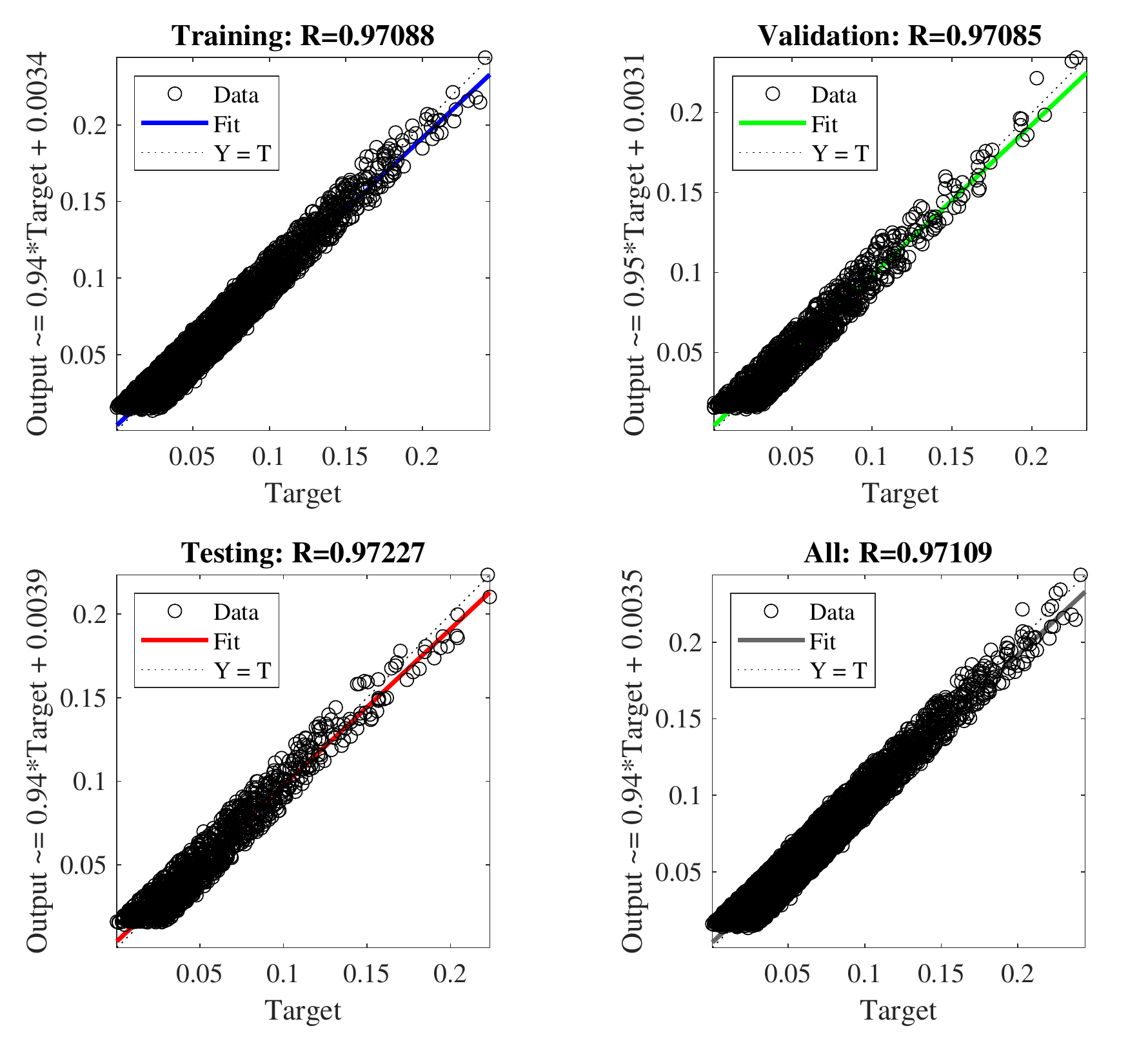}
\label{fig:61}
}
\subfigure[Correction effect under medium turbulent channel.]{
\includegraphics[width=0.48\linewidth]{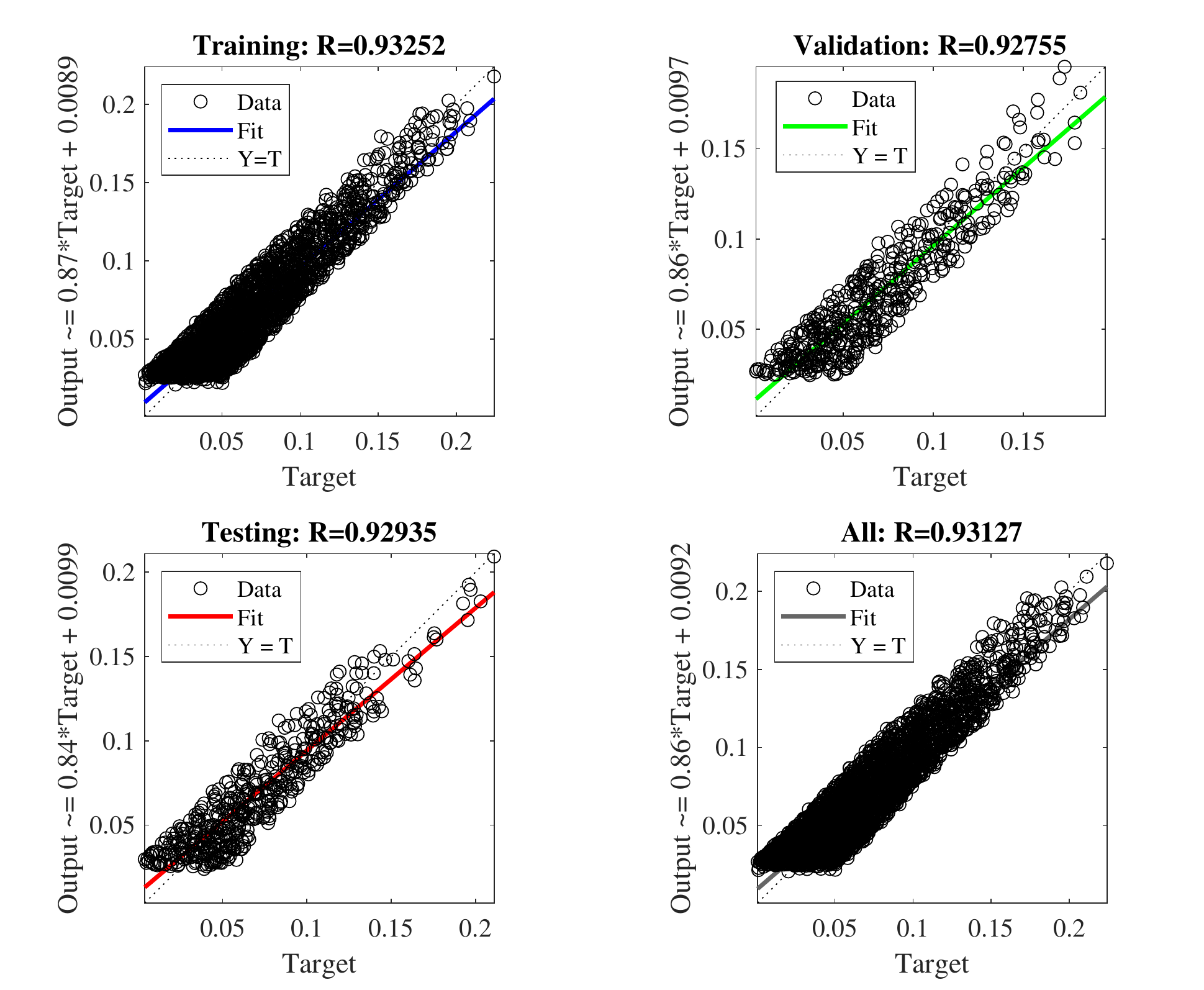}
\label{fig:62}
}
\subfigure[Correction effect under weak turbulent channel.]{
\includegraphics[width=0.43\linewidth]{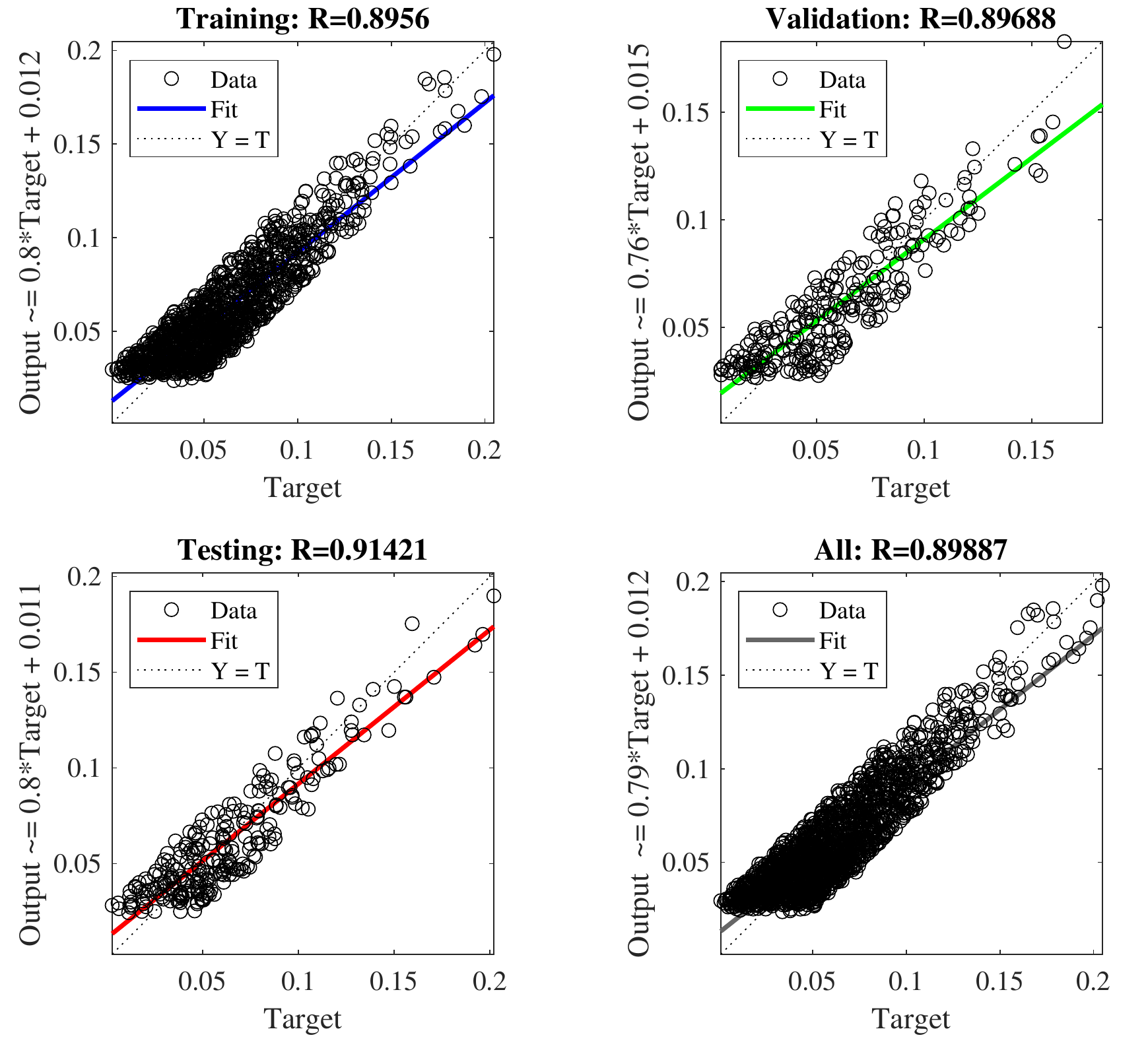}
\label{fig:63}
}
\subfigure[Correction effect without classification.]{
\includegraphics[width=0.48\linewidth]{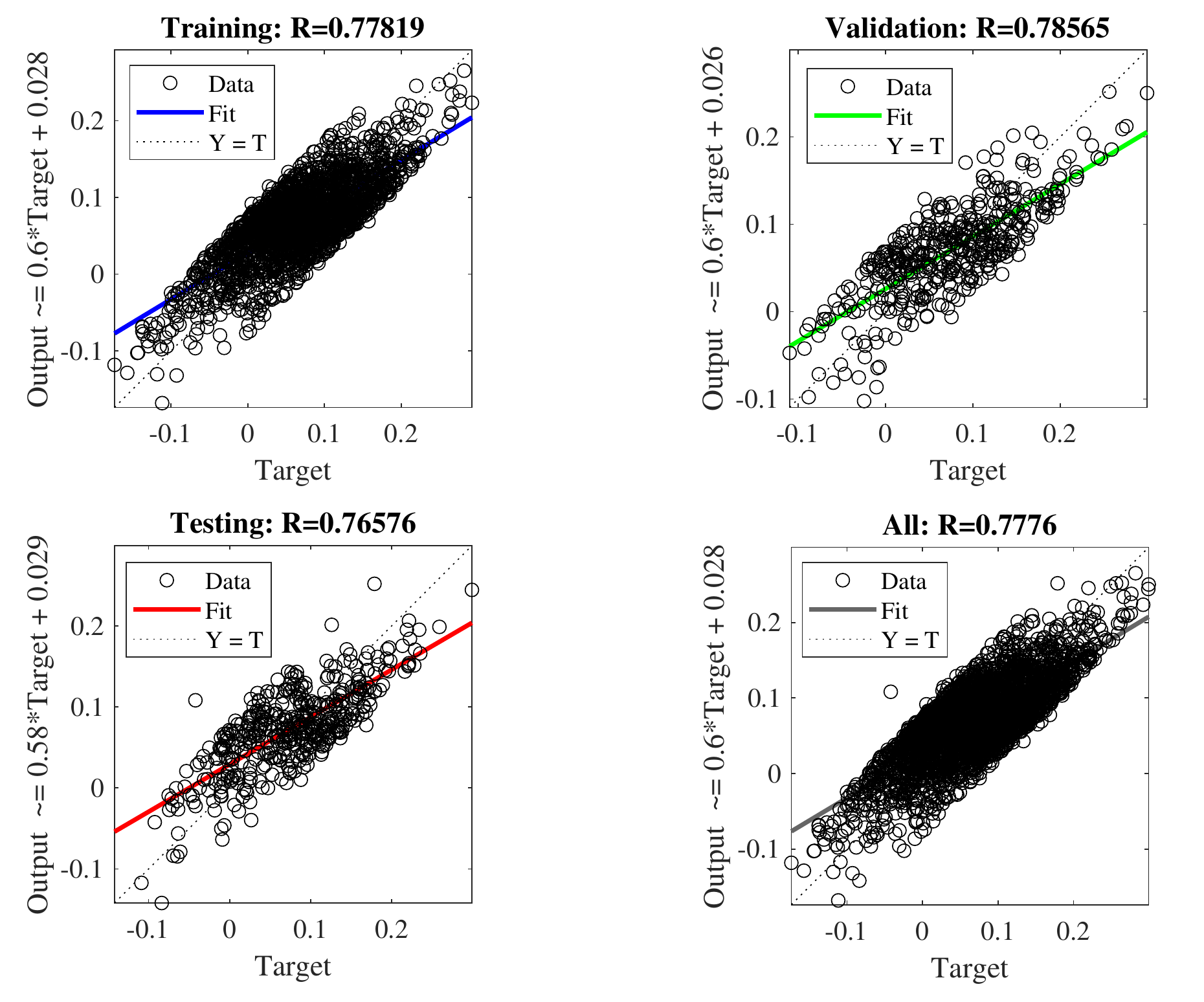}
\label{fig:64}
}
\caption{\textbf{Equalization correction via neural network and KNN.} (Figs. \ref{fig:61} - \ref{fig:63} describe the equalization correction under each turbulence,  Fig. \ref{fig:64} describes the equalization correction without classification. The slopes of blue, green, red, and black lines represent the degree of fit in the training set, validation set, test set, and the entire data set, respectively. The slope of the dotted line represents the best fit.)}
\label{fig:6}
\end{center}
\end{figure*}

Third, for different variable classes, the correction coefficients are obtained by different neural networks. It can be seen that when the variables are classified, the corrected results of the neural network under different turbulence channels are excellent, maintained at more than 90\% (Figs. \ref{fig:61} - \ref{fig:63}). However, if the classification algorithm is not employed and the prediction results of all variables are directly corrected, the prediction results are reduced, about 70\%  and more (Fig. \ref{fig:64}), which also proves the importance of early classification.

\begin{figure}[!htbp]\center
\centering
\resizebox{10cm}{!}{
\includegraphics{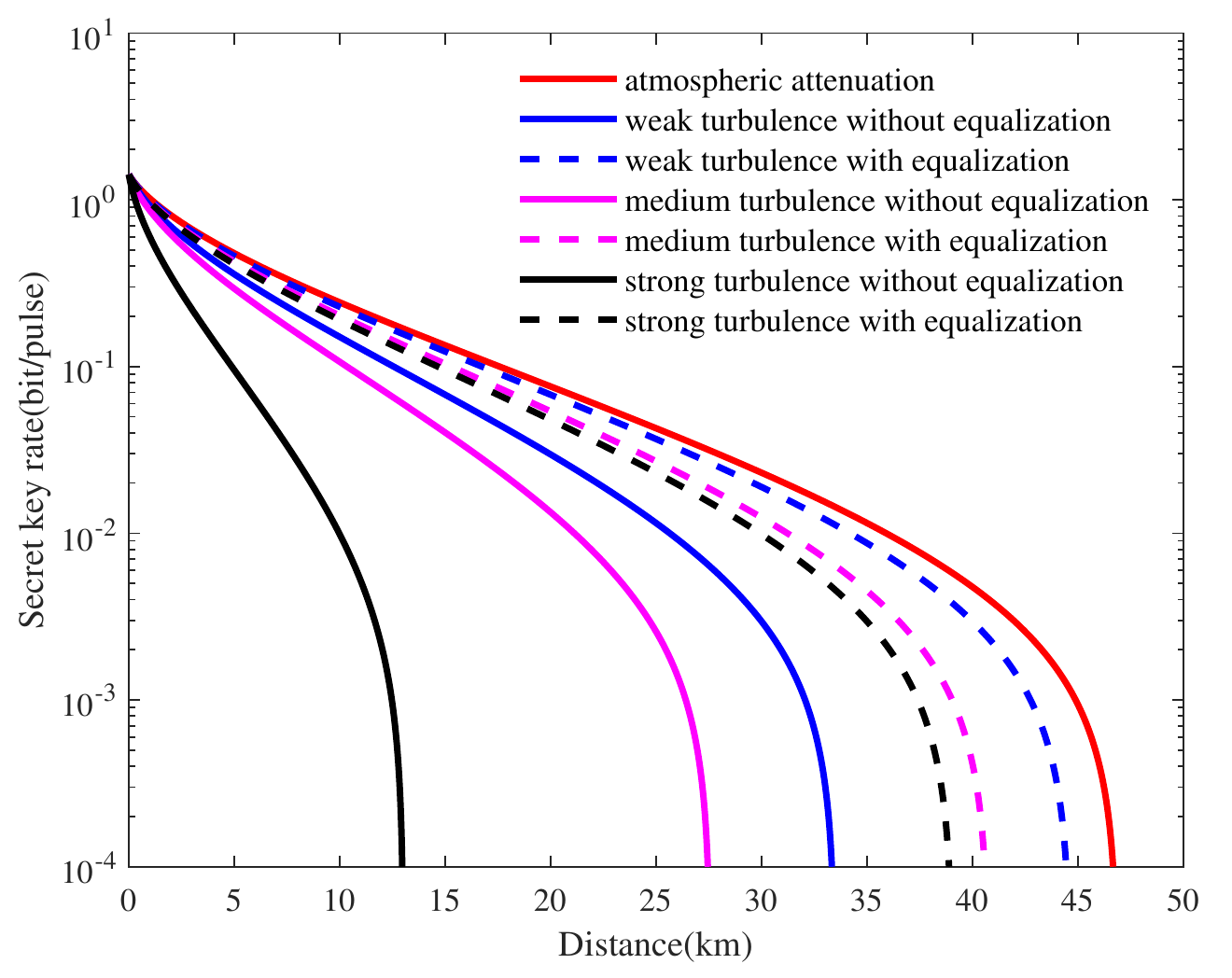}}
\caption{\textbf{Effect of equalization on system performance under different turbulence conditions.} (The solid red line represents the performance of the system considering only atmospheric attenuation. The blue lines, rose-red lines, and black lines represent the system performance under weak turbulence, medium turbulence, and strong turbulence, respectively. Among them, the solid lines and the dotted lines are the cases without equalization and with equalization, respectively.) }
\label{fig:7}
\end{figure}

The system performance is shown as Fig. \ref{fig:7}. It can be seen that the atmospheric attenuation model has the least impact on the system performance,  followed by weak turbulence and medium turbulence, and strong turbulence has the greatest influence.
The scheme is more universal and systematic with the assistance of a classification algorithm. Among them, the performance improvement of the system under medium-to-strong turbulence is the most obvious.

\section{Conclusion}
This work shows the excess noise suppression and system performance enhancement based on equalization in a GMCS CVQKD experiment with a 10km fiber link through the pilot tone and neural network. Its results show that the correction effect of equalization is excellent, which can reduce excess noise to the theoretical value, and improve the system performance. For the more complex volatility of the free-space channel, the KNN classification algorithm is added, and the scheme can be achieved more efficiently. For different classes of received variables, the trained correction coefficients are more targeted, which can cope with various turbulence, thus, this scheme fully considers the characteristics of the large change range and fast change speed of the free-space channel.  Compared with the unclassified equalization scheme, the classified scheme has a better correction effect, especially in medium-to-strong turbulence. Compared with other schemes, it focuses on the source of excess noise, fundamentally solves the problem of superfluous excess noise and boosts system performance, and omits phase tracking and compensation, reducing the hardware complexity and relieving the pressure on the system post-processing. It can be implemented on different links, which lays a foundation for the implementation of large-scale CVQKD in the future.

%
%
%

\section*{Acknowledgements}
This work is supported by the National Natural Science Foundation of China (Grant No. 62071381), Shaanxi Provincial Key R\&D Program General Project (2022GY-023), and ISN 23rd Open Project (ISN23-06) supported by the State Key Laboratory of Integrated Services Networks (Xidian University).

\section*{Appendix: The transmission-induced distortions}
As an indispensable component of the global quantum communication networks, the free-space CVQKD is drawing much more attention.
The atmospheric channel is an anisotropic time-varying medium composed of a variety of gas molecules and suspended particles, and its effect on the beam mainly includes the beam extinction caused by the absorption and scattering of atmospheric molecules, also called the atmospheric attenuation effect, and the random fluctuations in signal phase and amplitude caused by changes in atmospheric density and humidity caused by the motion of atmospheric molecules also called the atmospheric turbulence effect.

When beams pass through the atmospheric turbulence channel, they may deviate from the established transmission direction during propagation, resulting in beam wandering and broadening. The interference of refracted beam of different intensities in the receiving aperture section brings about the loss of phase, resulting in deformation and scintillation. These phenomena can be described by the elliptical beam model \cite{tuoyuan}. According to sub-channel theory \cite{sub}, it is assumed that in a sub-channel $i$-th with a transmittance of $T_{i}$ and an excess noise of $\varepsilon_{i} $, Alice modulation variable is $X^{i} =\left \{ x_{1}, x_{2},...,x_{N} \right \}$ and Bob measurement variable is $Y^{i} =\left \{ y_{1}, y_{2},...,y_{N} \right \} $, and the correlation data of the two $\left \{ \left ( x_{j}^{i} ,  y_{j}^{i}  \right )_{j=1, 2, ..., N}^{i=1, 2, ..., M}   \right \}$ ($\mathit{M} $ is the number of sub-channel, and $\mathit{N}$  is the number of quantum state transmitted in each sub-channel) meets the following relationship
\begin{eqnarray}
y_{j}^{i} =t_{i}x_{j}^{i} +z_{i},
\label{liner}
\end{eqnarray}
where $t_{i}=\sqrt{\eta T_{i} }$, $\eta$ is detection efficiency. The variance of Gaussian noise $z_{i}$ is $\sigma _{i}^{2}  =N_{0i}+ \eta T_{i}\varepsilon _{i}+\nu _{el}  $, $N_{0i}$ represents the shot noise in $i$-th channel.

According to the GMCS protocol, Alice modulates the information on the amplitude $A$ and phase $\varphi$ of pulses by  amplitude modulation and phase modulation, thus, $X^{i}$ can be further described as
\begin{eqnarray}
\begin{split}
X^{i}  =  A^{i} cos\varphi^{i}, 
\end{split}
\end{eqnarray}
and the Gaussian coherent state passes through the linear channel model of Eq. (\ref{liner}), $Y^{i}$ can be described as
\begin{eqnarray}
\begin{split}
Y^{i}  =  \sqrt{\eta } A_{\alpha }^{i} A^{i}cos(\varphi^{i}+\Delta\varphi^{i} )+A_{n}^{i} ,
\end{split}
\end{eqnarray}
where $A_{\alpha }^{i}$ and $\Delta\varphi^{i}$ represent the amplitude attenuation parameter and phase fluctuation caused by free-space quantum channel propagation. $A_{n}^{i}$ represents the amplitude of noise.

According to the Maximum likelihood estimation, parameters are estimated as
\begin{eqnarray}
\begin{split}
\hat{t_{i} } =\frac{\sum_{j=1}^{m}x_{j}^{i} y_{j}^{i} }{\sum_{j=1}^{m} x_{j}^{i2}   } ,
\hat{\sigma } _{i} =\frac{1}{m}\sum_{j=1}^{m}  (y_{j}^{i}-\hat{t }_{i}x_{j}^{i}),
\label{PE}
\end{split}
\end{eqnarray}
with the confidence interval
\begin{eqnarray}
\begin{split}
\bigtriangleup t_{i}=z_{\varepsilon _{PE}/2 } \sqrt{\frac{\hat{\sigma } _{i} }{mV_{A} } } ,
\bigtriangleup \hat{\sigma } _{i} =z_{\varepsilon _{PE}/2 }\frac{\hat{\sigma } _{i}\sqrt{2} }{\sqrt{m} },
\end{split}
\end{eqnarray}
where $m$ is the number of signals for parameter estimation, and $V_{A}$ is modulated variance.
When considering the effects of channel fluctuations to the system, the Eq. (\ref{PE}) is rewritten as
\begin{eqnarray}
\begin{split}
\label{eq1}
\hat{t} _{i}&=\frac{E\left [ \sqrt{\eta }A_{\alpha }^{i}(A^{i})^2 cos\varphi^{i} cos(\varphi^{i} +\Delta\varphi^{i} )   \right ] }{\hat{V}_{A}  }=\sqrt{\eta }(E\left [A_{\alpha}^{i} cos\Delta\varphi^{i}  \right ]-E\left [A_{\alpha}^{i} cos\Delta\varphi^{i}  \right ]  ),\\
\hat{\sigma } _{i}^{2}&=E\left [ (Y^{i})^2  \right ] -2\hat{t} _{i} E\left [ X^{i}Y^{i} \right ] +\hat{t}_{i}^{2} E\left [( X^{i})^2 \right ]  \\
&=\hat{V} _{A} \left \{ \eta \left ( E\left [ (A_{\alpha }^{i})^2 cos^{2}\Delta \varphi^{i}   \right ] + E\left [ (A_{\alpha }^{i})^{2} sin^{2}\Delta \varphi^{i}  \right ]\right.\right.\left.-2E\left [A_{\alpha }^{i}sin \Delta \varphi^{i}  \right ]E\left [A_{\alpha }^{i}cos \Delta \varphi^{i}  \right ]  \right ) \\
&-2\hat{t}_{i} \sqrt{\eta }(E\left [A_{\alpha }^{i}cos \Delta \varphi ^{i} \right ]\left. -E\left [A_{\alpha }^{i}sin \Delta \varphi^{i}  \right ]) + \hat{t} _{i}^{2}  \right \}+E\left [(A_{\alpha }^{i})^{2}  \right ] .
\end{split}
\end{eqnarray}

\textbf{Beam extinction.} Under the atmospheric turbulent channel, according to the elliptical model, the beam wandering causes a random displacement between the beam and the receiving plane, and the amplitude attenuation of the transmitted beam depends on the beam profile at the moment of reception and the displacement of its centroid relative to the center of the receiving aperture. In general, under weak turbulence conditions, the amplitude attenuation caused by atmospheric turbulence is described by a log-normal distribution. Under medium-strong turbulence conditions, the probability distribution function of amplitude attenuation caused by atmospheric turbulence is described by Gamma-Gamma distribution \cite{gama},
\begin{center}
\begin{eqnarray}
\begin{split}
P_{w} (I)=\frac{1}{I\sqrt{2\pi \sigma _{I}^{2}(\mathbf{r} ,L) } } \mathrm{exp} \left \{ -\frac{\left [\mathrm{ ln}(1/\left \langle I(\mathbf{r} ,L) \right \rangle ) +\sigma _{I}^{2}(\mathbf{r} ,L)/2  \right ]^{2}  }{2\sigma _{I}^{2}(\mathbf{r} ,L)}  \right \} ,\\
P_{m-s} (I)=\frac{2(\alpha \beta )^{(\alpha+ \beta)/2} }{\Gamma (\alpha )\Gamma (\beta )} I^{\frac{(\alpha+ \beta)}{2}-1  }J_{\alpha- \beta}(2\sqrt{\alpha \beta I} ),
\end{split}
\end{eqnarray}
\end{center}
where $\mathbf{r}$ represents the relative position between the centroid of the signal beam and the center of the receiving aperture, $\sigma _{I}^{2}(\mathbf{r} ,L)$ represents scintillation index, $\Gamma (\bullet )$ represents the Gamma function, $J_{(\alpha-\beta)}(\bullet) $ is the second type of correction Bezier function. The effective numbers of large-scale turbulence and small-scale turbulence during the scattering process is represented as $\alpha =\left [\mathrm{exp} (\sigma _{\mathrm{lnx} }^{2} )-1  \right ]^{-1}  $  and $\beta  =\left [\mathrm{exp} (\sigma _{\mathrm{lny} }^{2} )-1  \right ]^{-1}$ , respectively.

\textbf{Phase fluctuation.} Generally speaking, it is assumed that the phase fluctuations caused by turbulent atmospheres follow a Gaussian distribution with variance $\sigma _{\Delta \varphi }^{2} $. The probability density function of $\Delta \varphi$ can be expressed as
\begin{eqnarray}
\rho (\Delta \varphi)=\frac{1}{\sqrt{2\pi} \sigma _{\Delta \varphi } } \mathrm{exp}\Big (-\frac{({\Delta \varphi }^{2})}{2\sigma _{\Delta \varphi }^{2} } \Big).
\end{eqnarray}
Thus, the feature function $M(\omega )$ under the Fourier transform of $\rho (\Delta \varphi)$ is
\begin{eqnarray}
M(\omega )=\mathrm{exp} \left ( -\frac{\omega ^{2}\sigma _{\Delta \varphi }^{2}  }{2}  \right ) ,
\end{eqnarray}
and  $\sigma _{\Delta \varphi }^{2}$ can be described by Zernike polynomial as \cite{150}
\begin{eqnarray}
\sigma _{\Delta \varphi }^{2}  =  C_{J}\left ( \frac{2a}{d_{0} }  \right )  ^{2}, 
\end{eqnarray}
where $a$ is receive aperture radius, $d_{0}$ \cite{151} is used to describe the wavefront coherent diameter of the spatial correlation of phase fluctuations within the receiving plane, and $C_{J}$ is determined by Zernike polynomial \cite{152,153}. Each part in Eq. (\ref{eq1}) is described as

\begin{eqnarray}
\begin{aligned}
E\left [ A_{\alpha }^{i} cos\Delta \varphi^{i}  \right ]  &=   \frac{M(1)+M(-1)}{2} \int _{\mathfrak{B}} \sqrt{I(\mathbf{r_{i}} ,L)}p(I(\mathbf{r_{i}} ,L))d\mathbf{r_{i}  }\\
&=  \text{exp}\Big(-\frac{(\sigma _{\Delta \varphi }^{i})^{2} }{2}\Big )\int _{\mathfrak{B} } \sqrt{I(\mathbf{r_{i}} ,L)}p(I(\mathbf{r_{i}} ,L))d\mathbf{r_{i}  },\\
E\left [ A_{\alpha }^{i} sin\Delta \varphi ^{i} \right ] &=  \frac{-j[M(1)+M(-1)]}{2} \int _{\mathfrak{B} } \sqrt{I(\mathbf{r_{i}} ,L)}p(I(\mathbf{r_{i}} ,L))d\mathbf{r_{i}  },\\
E\left [ (A_{\alpha }^{i})^{2} cos^{2}\Delta \varphi^{i}  \right ]  &=  \frac{2+[M(2)+M(-2)]}{4} \int _{\mathfrak{B} } \sqrt{I(\mathbf{r_{i}} ,L)}p(I(\mathbf{r_{i}} ,L))d\mathbf{r_{i}  } \\
&=\frac{1+\mathrm{exp\Big(-(\sigma _{\Delta \varphi }^{i})^{2} \Big)} }{2} \int _{\mathfrak{B} } \sqrt{I(\mathbf{r_{i}} ,L)}p(I(\mathbf{r_{i}} ,L))d\mathbf{r_{i}  },\\
E\left [ (A_{\alpha }^{i})^{2} sin^{2}\Delta \varphi^{i}  \right ]  &= \frac{2-[M(2)-M(-2)]}{2} \int _{\mathfrak{B} } \sqrt{I(\mathbf{r_{i}} ,L)}p(I(\mathbf{r_{i}} ,L))d\mathbf{r_{i}  }\\
&= \frac{1-\mathrm{exp\Big(-(\sigma _{\Delta \varphi }^{i})^{2} \Big)} }{2} \int _{\mathfrak{B} } \sqrt{I(\mathbf{r_{i}} ,L)}p(I(\mathbf{r_{i}} ,L))d\mathbf{r_{i}  }.
\end{aligned}
\end{eqnarray}
Thus, Eq. (\ref{eq1}) can be smplified to 
\begin{eqnarray}
\begin{split}
\hat{t} _{i}&=\sqrt{\eta }  (E\left [ A_{\alpha }^{i} cos\Delta \varphi^{i}  \right ] -E\left [ A_{\alpha }^{i} sin\Delta \varphi^{i}  \right ])=\mathrm{exp}\Big(-\frac{(\sigma _{\Delta \varphi }^{i})^{2} }{2}\Big ) \int _{\mathfrak{B} } \sqrt{I(\mathbf{r_{i }} ,L)}p(I(\mathbf{r_{i}} ,L))d\mathbf{r_{i}  },\\
\sigma _{i}^{2}&=\hat{V}_{A}\eta    \left \{ E[(A_{\alpha }^{i})^{2}cos^{2}\Delta \varphi^{i}  ]+E[(A_{\alpha }^{i})^{2}sin^{2}\Delta \varphi^{i} ] \right. \left.-(E[A_{\alpha }^{i}cos\Delta \varphi^{i}])^2 \right \}+E[(A_{\alpha }^{i})^{2}] .
\end{split}
\end{eqnarray}
Thus, for sub-channel $i_{th}$, 
\begin{eqnarray}
\begin{split}
\hat{T} _{i}&=\frac{\hat{t} _{i}}{\eta } =\frac{\left [ \mathrm{exp}-\Big(\sigma _{\Delta \varphi }^{i})^{2}\Big ) \int _{\mathfrak{B} } \sqrt{I(\mathbf{r_{i }} ,L)}p(I(\mathbf{r_{i}} ,L))d\mathbf{r_{i}  }  \right ]^2 }{\eta },\\
\hat{\varepsilon} _{i}& =\frac{\hat{\sigma}_{i}^{2}-N_{0i}-\nu _{el}     }{\hat{t} _{i}} +\hat{V} _{A} \left \{ \frac{E\left [ (A_{\alpha }^{i})^{2} cos^{2}\Delta \varphi^{i}  \right ] }{(E\left [ A_{\alpha }^{i} cos\Delta \varphi ^{i} \right ])^{2}  }+\frac{E\left [ (A_{\alpha }^{i})^{2} sin^{2}\Delta \varphi^{i}  \right ] }{(E\left [ A_{\alpha }^{i} cos\Delta \varphi^{i}  \right ])^{2} }-1   \right \}.
\end{split}
\end{eqnarray}
Taking all subchannels into consideration, the estimated values of atmospheric channel transmittance $\left \langle \hat{T}\right \rangle $ and excess noise $ \hat{\varepsilon} $ are
\begin{eqnarray}
\begin{split}
\left \langle \hat{T} \right \rangle &=\sum_{i=1}^{M}p_{i}T_{i} =  \frac{1}{\eta } \sum_{i=1}^{M} p_{i} \left [ \mathrm{exp}\Big(-(\sigma _{\Delta \varphi ^{i}) }^{2}\Big ) \int _{\mathfrak{B} } \sqrt{I(\mathbf{r_{i }} ,L)}p(I(\mathbf{r_{i}} ,L))d\mathbf{r_{i}  }  \right ]^2,\\
 \hat{\varepsilon  }& =\sum_{i=1}^{M} p_{i}\varepsilon _{i}=\sum_{i=1}^{M} p_{i}\Bigg(  \frac{\hat{\sigma}_{i}^{2}-N_{0i}-\nu _{el}     }{\hat{t}_{i} }+\hat{V} _{A} \left \{ \frac{E\left [ (A_{\alpha }^{i})^{2} cos^{2}\Delta \varphi^{i} \right ] }{(E\left [ A_{\alpha }^{i} cos\Delta \varphi^{i}  \right ])^{2}  }+\frac{E\left [ A_{\alpha }^{i} sin^{2}\Delta \varphi ^{i} \right ] }{(E\left [ A_{\alpha }^{i} cos\Delta \varphi^{i} \right ])^{2} }-1   \right \}\Bigg).
\end{split}
\end{eqnarray}
Thus, the asymptotic secret key rate of the free-space CVQKD system in reverse reconciliation is expressed as
\begin{eqnarray}
K(\left \langle \hat{T } \right \rangle,\hat{\varepsilon}  )=(1-\text{FER}) \left [ \beta _{R} I_{AB}(\left \langle \hat{T } \right \rangle,\hat{\varepsilon} )-\chi _{BE} (\left \langle \hat{T } \right \rangle,\hat{\varepsilon})-\bigtriangleup (n)\right ],
\end{eqnarray}
where $\text{FER} \in \left [ 0,1 \right ] $, representing the frame-error rate, and $\beta _{R} \in \left [ 0,1 \right ] $, representing the efficiency of information reconciliation. $ I_{AB}$ means mutual information between Alice and Bob, $\chi _{BE}$ means the maximum amount of information that Eve can steal from Bob limited by the Holevo bound, and $\bigtriangleup (n)$ is related to the security parameter of the privacy amplification.
\end{document}